\documentclass[12pt]{article}
\pdfoutput=1
\usepackage{myjheppub}
\usepackage{graphicx,booktabs,latexsym,amssymb,amsmath,bm,epsfig,psfrag,float}
\hypersetup{pdfnewwindow=true}

\preprint{
\begin{flushright}	

	TUM-HEP-1075/17,\\
	IPPP/17/4
\end{flushright}}

\title{Associated production of a top pair and a Z boson at the LHC to NNLL accuracy}

\author[a,]{Alessandro Broggio,}
\author[b,c]{Andrea Ferroglia,}
\author[b,c]{Giovanni Ossola,}
\author[d]{Ben D. Pecjak,}
\author[b,c]{and Ray D. Sameshima}

\affiliation[a]{Physik Department T31, Technische Universit\"at M\"unchen,
James Franck-Stra{\ss}e 1, D-85748 Garching, Germany}
\emailAdd{alessandro.broggio@tum.de}

\affiliation[b]{Physics Department, New York City College of Technology, The City
	University of New York, 
	300 Jay Street, Brooklyn, NY 11201 USA}

\affiliation[c]{The Graduate School and University Center,
The City University of New York, 365 Fifth Avenue,
New York, NY 10016  USA}
\emailAdd{aferroglia@citytech.cuny.edu}
\emailAdd{gossola@citytech.cuny.edu}

\affiliation[d]{Institute for Particle Physics Phenomenology, Ogden Centre for Fundamental Physics,
	Department of Physics, University of Durham, Science Laboratories,
	South Rd, Durham DH1 3LE, United Kingdom}
\emailAdd{ben.pecjak@durham.ac.uk}

\emailAdd{rsameshima@gradcenter.cuny.edu}

\abstract{We study the resummation of soft gluon emission corrections
  to the production of a top-antitop pair in association with a Z
  boson at the Large Hadron Collider to next-to-next-to-leading logarithmic accuracy. By means of an in-house parton level Monte Carlo code we evaluate the resummation formula for the total cross section and several differential distributions at a center-of-mass energy of $13$~TeV, and we match these calculations to next-to-leading order results.}

\begin{document}
\newcommand{\Red}[1]{\textcolor{red}{#1}}
\newcommand{\Green}[1]{\textcolor{green}{#1}}
\newcommand{\Blue}[1]{\textcolor{blue}{#1}}
\newcommand{\Cyan}[1]{\textcolor{cyan}{#1}}
\newcommand{\Magenta}[1]{\textcolor{magenta}{#1}}
\newcommand{\alert}[1]{{\bf \color{red}[{#1}]}}

\newcommand{\be}{\begin{equation}}
\newcommand{\ee}{\end{equation}}

\newcommand{\nn}{\nonumber}
\def\ff{f\hspace{-0.3cm}f}
\def\mgamc{{{\tt \small MG5\_aMC}}}

\maketitle


\section{Introduction}
\label{sec:intro}

The associated production of a top pair and a $Z$ or $W$ boson are the
two processes with the heaviest final states measured to date at the
Large Hadron Collider (LHC).  The total cross section for these
processes was measured during Run I
\cite{Khachatryan:2015sha,Aad:2015eua}, and preliminary measurements
at a center-of-mass energy of 13~TeV are also available
\cite{ATLAS13,CMS:2016dui}. The $t\bar{t} Z$ production process is
particularly interesting because it allows one to study the coupling
of the $Z$ boson with the top quark. This measurement further tests
the Standard Model (SM) of particle physics and probes several Beyond
the SM scenarios that predict changes to this coupling with respect to
the SM. In addition, these production processes lead to high
multiplicity final states which are background in the search for new
heavy states decaying via long chains, such as dark matter candidates.

Given their importance for phenomenological studies, next-to-leading-order (NLO) QCD and electroweak corrections to the associated production of a top pair and a massive vector boson were studied by several groups~\cite{Lazopoulos:2007bv,Lazopoulos:2008de,Garzelli:2012bn,Campbell:2012dh,Maltoni:2014zpa,Maltoni:2015ena,Frixione:2014qaa,Frixione:2015zaa}. A full calculation of the QCD corrections to next-to-next-to leading order (NNLO) accuracy would be desirable but it is extremely difficult even with the most up to date techniques for the calculations of higher order corrections. However, the associated production of a top pair and a heavy colorless boson is a multiscale process which is expected to receive potentially large corrections arising from soft gluon emission.  The resummation of these effects to next-to-next-to leading logarithmic (NNLL) accuracy can be carried out by exploiting the factorization properties of the partonic cross section in the soft limit (which can be studied with effective field theory methods\footnote{For an introduction see \cite{Becher:2014oda}.}) and by subsequently  employing renormalization group improved perturbation theory techniques. In the case of the associated production of a top pair and a Higgs boson the resummation formula in the soft emission limit  was discussed in \cite{Broggio:2015lya}, and results for the total cross section and several differential distributions at NLO+NNLL accuracy were presented in \cite{Broggio:2016lfj}.
Studies of the associated production of a top quark pair and a $W$ boson to NLO+NNLL accuracy  can be found in \cite{Broggio:2016zgg}, where the resummation was carried out in Mellin moment space as in   \cite{Broggio:2016lfj}, and in \cite{Li:2014ula}, where the resummation was instead carried out in momentum space.

The results of  \cite{Broggio:2016lfj} and \cite{Broggio:2016zgg} were obtained by means  of  an in-house parton level Monte Carlo code for the numerical evaluation of the resummation formula.
The output of this code was then matched to complete  NLO calculations obtained by employing 
\verb|MadGraph5_aMC@NLO| \cite{Alwall:2014hca} (which we indicate with \mgamc~in the following). Building on the results of those two papers, in this work we obtain a resummation formula for the associated production of  $t \bar{t} Z$ final state, and we evaluate it to NNLL accuracy by means of dedicated parton level Monte Carlo code. We match our results for the total cross section and differential distributions to NLO calculations in order to obtain predictions at
NLO+NNLL accuracy.

The paper is organized as follows: In Section~\ref{sec:outline} we introduce some basic notation and we briefly summarize the main steps in our calculations. For a more technical discussion of the methods employed in this paper, we refer the reader to the detailed descriptions provided in~\cite{Broggio:2015lya,Broggio:2016lfj,Broggio:2016zgg}. In Section~\ref{sec:numbers} we present predictions at NLO+NNLL accuracy for the total cross section as well as for several differential distributions. Finally, we draw our conclusions in Section~\ref{sec:conclusions}.

\section{Outline of the Calculation}
\label{sec:outline}

The associated production of a top quark pair and a $Z$ boson receives contributions from the partonic process 
\begin{align}
i(p_1) +  j(p_2) \longrightarrow t(p_3) +\bar{t} (p_4) + Z(p_5)  + X \, , 
\label{eq:partproc}
\end{align}
where $ij \in \{q \bar{q}, \bar{q}q, gg\}$ at lowest order in QCD, and
$X$ indicates the unobserved partonic final-state radiation. The two 
Mandelstam invariants which are relevant for our discussion are 
\begin{align}
\hat{s} = (p_1+p_2)^2  = 2 p_1 \cdot p_2 \, , \quad \text{and} \quad
M^2 = \left(p_3 + p_4 + p_5 \right)^2 \, .  \label{eq:mandelstam}
\end{align}
The soft or partonic threshold limit is defined as the kinematic region in which
$z \equiv M^2/\hat{s} \rightarrow 1$.
In this region, the final state radiation indicated by $X$ in (\ref{eq:partproc}) can only be soft.

The factorization formula for the QCD cross section in the partonic 
threshold limit is the same as the one derived in 
\cite{Broggio:2015lya} for $t \bar{t} H$ production, up to the straightforward replacement of the Higgs boson with a $Z$ boson:
\begin{align}
\sigma \left(s,m_t,m_Z \right) =& \frac{1}{2 s} \int_{\tau_{\text{min}}}^{1} \!\!\! d \tau \int^1_{\tau} \frac{dz}{\sqrt{z}} \sum_{ij} \ff_{ij} \left( \frac{\tau}{z}, \mu \right)\nonumber \\
& \times \int d\text{PS}_{t\bar{t}Z} \mbox{Tr}\left[\mathbf{H}_{ij}\left(\{p\},\mu\right) \mathbf{S}_{ij}\left(\frac{M (1-z)}{\sqrt{z}},\{p\},\mu\right)  \right] \, .
\label{eq:factorization}
\end{align}
We indicated with $s$ the square of the hadronic 
center-of-mass energy and we defined 
$\tau_{\text{min}} = \left(2 m_t + m_Z\right)^2/s$ and $\tau = M^2/s$.
The notation adopted for the channel dependent hard functions
$\mathbf{H}$, soft functions $ \mathbf{S}$, and luminosity functions
$\ff$, as well as for the final-state phase-space integration measure,
is the same one used in \cite{Broggio:2016lfj, Broggio:2016zgg} and
we refer the reader to these papers for more details.  Similarly to LO,
the only subprocesses to be considered in the soft limit are those
labeled by indices $ij\in\{q\bar{q},\bar{q}q,gg\}$. The hard and soft
functions are two-by-two matrices in color space for $q\bar{q}$-initiated (quark
annihilation) processes, and three-by-three matrices in color space for
$gg$-initiated (gluon fusion) processes.  Contributions from other
production channels such as $\bar{q}g$ and $qg$ (collectively referred
to as ``quark-gluon'' or simply ``$qg$'' channel in what follows) are
subleading in the soft limit.
The hard functions satisfy renormalization group equations governed by
the channel dependent soft anomalous dimension matrices $\mathbf{\Gamma}^{ij}_H$.  These anomalous dimension matrices were derived in \cite{Ferroglia:2009ep,Ferroglia:2009ii}.

In order to carry out the resummation to NNLL accuracy, the hard functions, soft functions, and soft anomalous dimensions must be computed in fixed-order perturbation theory up to NLO in
$\alpha_s$. The NLO soft functions and soft anomalous dimensions are the same ones needed in the calculation of $t \bar{t} H$ and $t \bar{t} W^\pm$ to NNLL accuracy and can be found in 
\cite{Broggio:2015lya,Broggio:2016lfj,Broggio:2016zgg}. 
 The NLO hard functions are instead process dependent, receive contributions exclusively from the virtual corrections to the tree level amplitudes,  and were evaluated by customizing the one-loop provider {\tt Openloops} \cite{Cascioli:2011va}, which we used in combination with the tensor reduction library {\tt Collier} 
\cite{Denner:2002ii,Denner:2005nn,Denner:2010tr,Denner:2016kdg}. The NLO hard function have been cross-checked numerically by means of a customized version of {\tt GoSam}~\cite{Cullen:2011ac, Cullen:2014yla,Binoth:2008uq,Mastrolia:2010nb}, used in combination with the reduction provided by {\tt Ninja}~\cite{Mastrolia:2012bu,vanDeurzen:2013saa,Peraro:2014cba}.

In this paper we carry out the resummation in Mellin space, starting from the relation
\begin{align}
\sigma(s,m_t,m_Z) = & \frac{1}{2 s} \int_{\tau_{\text{min}}}^{1} \frac{d \tau}{\tau} \frac{1}{2 \pi i} \int_{c - i \infty}^{c + i \infty} dN \tau^{-N} \sum_{ij} \widetilde{\ff}_{ij}\left(N, \mu \right) \int d \text{PS}_{t \bar{t} Z} \, \widetilde{c}_{ij} \left(N,\mu\right) \, ,
\label{eq:Mellinfac}
\end{align}
where $\widetilde{\ff}_{ij}$ is the Mellin transform of the luminosity functions, and $\widetilde{c}$ is the Mellin transform of the product 
of the hard and soft function (see \cite{Broggio:2016zgg,Broggio:2016lfj} for details).
%
Since the soft limit $z \to 1$ corresponds to the limit $N \to \infty$
in Mellin space, we neglected terms suppressed by powers of $1/N$ in the integrand of
(\ref{eq:Mellinfac}). 

The hard and soft functions included in the hard scattering kernels $\widetilde{c}$ in (\ref{eq:Mellinfac}) can be evaluated
in fixed order perturbation theory at scales at which they are free
from large logarithms. We indicate these scales with $\mu_h$ and
$\mu_s$, respectively. Subsequently, by solving the renormalization group (RG) equations for
the hard and soft functions one can evolve the factor
$\widetilde{c}$ to the factorization scale $\mu_f$. Following this procedure one finds
\begin{align}
\widetilde{c}_{ij}(N,\mu_f) =  
\mbox{Tr} \Bigg[&\widetilde{\mathbf{U}}_{ij}(\!\bar{N},\{p\},\mu_f,\mu_h,\mu_s) \, \mathbf{H}_{ij}( \{p\},\mu_h) \, \widetilde{\mathbf{U}}_{ij}^{\dagger}(\!\bar{N},\{p\},\mu_f,\mu_h,\mu_s)
\nn \\
& \times \widetilde{\mathbf{s}}_{ij}\left(\ln\frac{M^2}{\bar{N}^2 \mu_s^2},\{p\},\mu_s\right)\Bigg] \,  ,
\label{eq:Mellinresum}
\end{align}
where $\bar{N}=N e^{\gamma_E}$.
Large logarithmic corrections depending on the ratio
of the scales $\mu_h$ and $\mu_s$ are resummed in the
channel-dependent matrix-valued evolution factors
$\widetilde{\mathbf{U}}$. The expression for the evolution factors is formally identical to the one found for $t \bar{t} W$ and $t \bar{t} H$ production and can be found for example in equation (3.7) of \cite{Broggio:2016zgg}.

The l.h.s of (\ref{eq:Mellinresum}) is formally independent of $\mu_h$ and $\mu_s$.
In practice however, one cannot evaluate the hard and soft functions at all orders in perturbation theory; this fact creates a residual dependence on the choice of the scales $\mu_h$
and $\mu_s$ in any numerical evaluation of $\widetilde{c}$. The hard and soft functions are free from large logarithms if one chooses $\mu_h \sim M$ and $\mu_s \sim
M/\bar{N}$. The choice of  a $N$-dependent value for  $\mu_s$ produces a branch cut for large values of $N$ in the hard scattering kernels $\widetilde{c}$,
whose existence is related to the Landau pole in $\alpha_s$. We deal with this issue by  choosing the integration path in the complex $N$ plane according to the
{\emph{Minimal Prescription}} (MP) introduced in \cite{Catani:1996yz}.

Finally, the parton luminosity functions in Mellin
space, which we need in the numerical evaluations, are constructed using techniques described in 
\cite{Bonvini:2012sh, Bonvini:2014joa}.

\section{Numerical Results}
\label{sec:numbers}

\begin{table}[t]
	\begin{center}
		\def\arraystretch{1.3}
		\begin{tabular}{|c|c||c|c|}
			\hline $M_W$ & $80.385$~GeV & $m_t$ & $173.2$~GeV\\ 
			\hline $M_Z$ &  $91.1876$~GeV & $m_H$ & $125$~GeV \\ 
			\hline $1/\alpha$ & $137.036$ & $\alpha_s \left(M_Z\right)$ & from MMHT 2014 PDFs \\ 
			\hline 
		\end{tabular} 
		\caption{Input parameters employed throughout the calculation. \label{tab:tabGmu}}
	\end{center}
\end{table}

The main purpose of this section is to present predictions for the
associated production of a top pair and a $Z$ boson to NLO+NNLL
accuracy. However, we also analyze systematically the relevance of
soft emission corrections and their resummation in relation to NLO
predictions for the various observables considered in the paper.  The
NNLL calculations are carried out by means of an in-house parton level
Monte Carlo code, while the NLO predictions are obtained by means of
\mgamc. All of the calculations discussed in this section are carried
out with the input parameters listed in
Table~\ref{tab:tabGmu}. Throughout the paper we employ MMHT 2014 PDFs
\cite{Harland-Lang:2014zoa}. In fixed order calculations, the order of
the PDFs matches the perturbative order of the calculation (i.e. LO
calculations are carried out with LO PDFs, NLO calculations employ NLO
PDFs, etc.). In matched calculations, we employed NLO PDFs for NLO+NLL
accuracy, and NNLO PDFs for NLO+NNLL accuracy.

For both the total cross section and several differential distributions, we consider six different types of predictions:
\begin{itemize}
	\item[\emph{i)}] {\bf NLO} calculations, obtained with \mgamc.
	\item[\emph{ii)}] {\bf Approximate NLO} calculations, obtained from the NLO expansion of the NNLL resummation formula. We check that for our choice of scales and input parameters approximate NLO calculations  provide a satisfactory approximation to the exact NLO calculation. The approximate NLO formulas obtained by expanding (\ref{eq:Mellinresum}) account for the single and double powers of $\ln N$ as well as $N$-independent terms but not for terms suppressed by inverse powers of $N$. $N$-independent terms depend on the Mandelstam
	variables, however we refer to them as ``constant" terms in what follows. 
	The approximate NLO formulas are obtained by setting $\mu_h=\mu_s=\mu_f$ in the NNLL version of (\ref{eq:Mellinresum}).
	Approximate NLO calculations are carried out with the in-house parton level Monte Carlo code which was developed specifically for this project.
	\item[\emph{iii)}] {\bf NLO+NLL} calculations, which are obtained by matching NLO results with resummed results at NLL accuracy obtained by means of the in-house Monte Carlo code.
	The results are matched according to the formula
	\begin{align}
	\sigma^{{\text{NLO+NLL}}}  =& 
	\sigma^{{\text{NLO}}}  + 
	\left[\sigma^{{\text{NLL}}}  -  \sigma^{{\text{NLL expanded to NLO}}} \right]  \, .
	\label{eq:NLOpNLLmatching_old}
	\end{align}
	The terms in the square brackets, which contribute to NLO and beyond, depend on the scales $\mu_s$ and $\mu_h$ in addition to the factorization scale $\mu_f$. Of course the dependence on $\mu_s$ and $\mu_h$ is formally of NNLL order; by varying these scales and the factorization scale in  (\ref{eq:NLOpNLLmatching_old}) one can estimate the size of the NNLL corrections.

	\item[\emph{iv)}] {\bf NLO+NNLL} predictions are obtained by evaluating the hard scattering kernels in (\ref{eq:Mellinresum}) to NNLL accuracy with the in-house Monte Carlo code and by matching the results to NLO calculations as follows:
	\begin{align}
	\sigma^{{\text{NLO+NNLL}}}  =&  \sigma^{{\text{NLO}}}
	+\left[ \sigma^{\text{NNLL}}- \sigma^{\text{approx. NLO}}\right]\,.
	\label{eq:NLOpNNLLmatching}
	\end{align}
	The terms in the squared brackets  in (\ref{eq:NLOpNNLLmatching}) contribute starting from NNLO and represent the NNLL corrections to be added to the NLO result.
	
	\item[\emph{v)}] {\bf Approximate NNLO} calculations are obtained by the NNLL resummation formula and  include all powers of $\ln N$ and part of the constant terms from a complete NNLO calculation. The approximate NNLO formulas employed in this paper are constructed as the ones employed in \cite{Broggio:2016zgg,Broggio:2016lfj} for $t \bar{t} W$ and $t \bar{t} H$ production. A detailed description of the constant terms which are included in the approximate NNLO formulas can be found in Section~4 of \cite{Broggio:2015lya}. Approximate NNLO formulas are evaluated with the in-house Monte Carlo code which we developed and they are matched to the NLO calculations as follows 
	  	\begin{align}
	  	\sigma^{\text{nNLO}} =\sigma^{\text{NLO}} 
	  	+\left[\sigma^{\text{approx. NNLO}} - \sigma^{\text{approx. NLO}}\right] \, ,
	  	\label{eq:NNLOmatching}
	  	\end{align}
	  	where we label the matched result ``nNLO'' for brevity. By construction nNLO predictions are independent from the hard and soft scales but they do have a residual N$^3$LO dependence on $\mu_f$. 
	  	
	\item[\emph{vi)}] {\bf NLO+NNLL expanded to NNLO}. Finally we consider a second way of expanding the NNLL resummation formula to NNLO. This approach differs from
	the approximate NNLO result used above by constant terms, which are formally of
	N$^3$LL accuracy. This approximation is defined by the relation
	\begin{align}
	\label{eq:NNLLexpanded}
	\left(\sigma^{\text{NLO+NNLL}}\right)_{\text{NNLO exp.}} &=
	\sigma^{\text{NLO}} +
	\left[\sigma^{\text{NNLL expanded to NNLO}}
	- \sigma^{\text{approx. NLO}}\right] \,.
	\end{align}
	The constant pieces in  (\ref{eq:NNLLexpanded}) contain explicit dependence on $\mu_h$ and $\mu_s$, in addition to that on $\mu_f$.  This dependence is formally an effect of N$^3$LL order. By comparing the predictions obtained from (\ref{eq:NNLLexpanded}) to the corresponding NLO+NNLL calculations one can see  the relative weight of  terms of N$^3$LO and higher in the NLO+NNLL calculations. If in the future a complete NNLO calculation for the $t \bar{t} Z$ production cross section were to become available, it would be possible to match it to the NNLL resummation formula by using precisely this kind of NNLO expansion of the NNLL resummation, as can be seen by replacing N~$\to$~NN in all of the superscripts in (\ref{eq:NLOpNLLmatching_old}).

\end{itemize}



\subsection{Scale choices}

\begin{figure}[tp]
	\begin{center}
		\includegraphics[width=10.5cm]{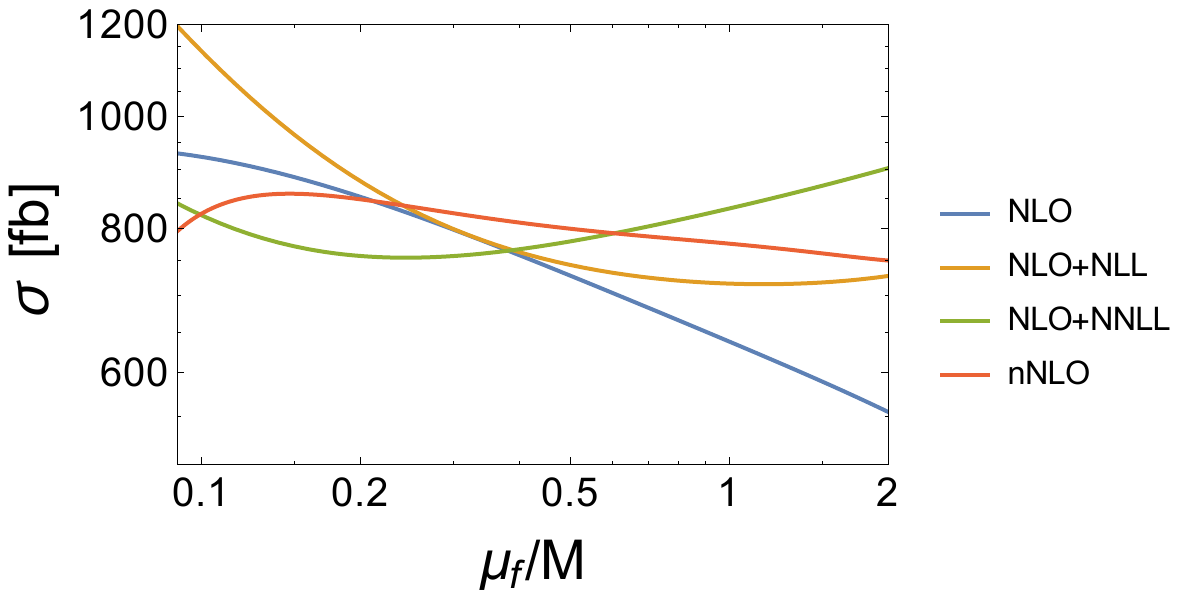} 
	\end{center}
	\caption{Factorization-scale dependence of the total $t\bar{t}Z$ production cross section 
		at the LHC with $\sqrt{s}=13$~TeV. The NLO and NLO+NLL curves are obtained 
		using MMHT 2014 NLO PDFs, while the NLO+NNLL and nNLO curves are obtained using  MMHT 2014 NNLO PDFs.
		\label{fig:scaledep}
	}
\end{figure}

Since any numerical evaluation of the resummed expression for the hard scattering kernels must be carried out by evaluating the factors in (\ref{eq:Mellinresum}) up to a certain order in perturbation theory, the resummed kernels $\widetilde{c}$ will show a residual dependence on the scales $\mu_s$ and $\mu_h$. In order to follow closely the approach adopted in ``direct QCD'' calculations \cite{Catani:1996yz, Bonvini:2012az,Bonvini:2014qga},   the standard choice which we adopt in this work for the hard and soft scales is  $\mu_{h,0} = M$ and $\mu_{s,0} =
M/ \bar{N}$ \cite{Ferroglia:2015ivv,Pecjak:2016nee,Broggio:2016zgg,Broggio:2016lfj}.

In addition, both fixed order and resummed calculations depend on the factorization scale $\mu_f$, which should be chosen in such a way that the logarithms of the scale ratio $\mu_f/M$
are not large \cite{Collins:1989gx}. 
It is therefore reasonable to choose a dynamical default value for the scale $\mu_f$ which is related to $M$. The dependence of the total $t \bar{t} Z$ production cross section on the ratio $\mu_f/M$ at the LHC with $\sqrt{s}=13$~TeV is shown in  Figure~\ref{fig:scaledep}. Each line corresponds to a different perturbative approximation, as indicated in the legend. Figure~\ref{fig:scaledep} shows that the NLO, NLO+NLL and NLO+NNLL curves intersect in the vicinity of $\mu_f/M = 0.5$ and differ significantly for $\mu_f/M \ll 0.5$ and for $\mu_f/M \gg 0.5$. Following this observation, the default value that we employ for the factorization scale is $\mu_{f,0} = M/2$.

The uncertainty related to the choice of the factorization scale in fixed order results is estimated as usual by varying this scale in the range $\mu_f \in [\mu_{f,0}/2, 2 \mu_{f,0}]$. Resummed results depend also on the hard and soft scales, consequently, the uncertainty of the resummed results is estimated by varying separately all the three scales around their default values
in the interval $\mu_i \in [\mu_{i,0}/2, 2 \mu_{i,0}]$ for $i \in \{s,f,h\}$. The scale uncertainty above (below) the central value of a resummed observable $O$, which can be the total cross section or the value of the differential cross section in a given bin, is determined as follows. First one evaluates the quantities 
\begin{align}
\Delta O_i^+= \text{max}\{O \left(\kappa_i=1/2\right),  O \left(\kappa_i=1\right), O \left(\kappa_i=2\right)\} - \bar{O} \, , \nn \\
\Delta O_i^-= \text{min}\{O \left(\kappa_i=1/2\right),  O \left(\kappa_i=1\right), O \left(\kappa_i=2\right)\} - \bar{O} \, , \label{eq:DeltaO} 
\end{align}
for $i \in \{s,f,h\}$. In (\ref{eq:DeltaO}) we defined $\kappa_i = \mu_i/ \mu_{i,0}$, and $\bar{O}$ indicates the observable evaluated at $\kappa_i =1$ for all $i$-s. The scale uncertainty above (below) $\bar{O}$ is then obtained by combining in quadrature $\Delta O_i^+$
($\Delta O_i^-$) for $i \in \{s,f,h\}$.

\subsection{Total cross section}
\label{sec:CS}

\begin{table}[t]
	\begin{center}
		\def\arraystretch{1.3}
		\begin{tabular}{|c|c|c|c|}
			\hline  order & PDF order & code & $\sigma$ [fb]\\ 
			\hline LO & LO & \mgamc & $ 521.4^{+165.4}_{-116.9} $ \\
			\hline \hline 
			\hline app. NLO & NLO & in-house MC & $ 737.7^{+38.5}_{-64.5} $ \\
			\hline NLO no $qg$ & NLO & \mgamc & $ 730.4^{+41.8}_{-64.9}$ \\
		         \hline NLO & NLO & \mgamc & $ 728.3^{+93.8}_{-90.3} $ \\
			\hline \hline NLO+NLL & NLO&  in-house MC  +\mgamc  & $742.0^{+90.1}_{-30.3} $ \\
		\hline NLO+NNLL  & NNLO & in-house MC +\mgamc  & $777.8^{+61.3}_{-65.2} $ \\
		\hline 
		\hline nNLO  & NNLO & in-house MC +\mgamc  & $798.7^{+36.2}_{-23.6} $ \\
		\hline 
			(NLO+NNLL)$_{\rm NNLO \, exp.}$  & NNLO & in-house MC +\mgamc  & $766.2^{+17.2}_{-50.1} $ \\	
								\hline 
		\end{tabular} 
		\caption{Total cross section for $t \bar{t} Z$ production at the
                  LHC with $\sqrt{s} = 13$~TeV and MMHT 2014 PDFs. The default value
                    of the factorization scale is $\mu_{f,0}=M/2$, and
                    the uncertainties are estimated through 
                    variations of this scale (and of the hard and soft  scales
                    $\mu_s$ and $\mu_h$ when applicable), as explained in the text.
\label{tab:CSHp13hM}}
	\end{center}
\end{table}

In this section we analyze the total cross section for the associated production of a top quark pair and a $Z$ boson at the LHC operating at a center-of-mass energy of $13$~TeV. The relevant results are collected in Table~\ref{tab:CSHp13hM}. We first compare the approximate NLO cross section, obtained by expanding the resummation formula to NLO (second row of  Table~\ref{tab:CSHp13hM}) with the complete NLO cross section (fourth row) and the NLO cross section without the contribution of the quark-gluon channel (third row).
The difference between the approximate NLO result and the NLO result without the $qg$ channel is due to terms in the quark annihilation and gluon fusion channels  which are subleading in the partonic threshold  limit. We see that the impact of these terms is around $1 \%$. The difference between these two results is therefore small in spite of the fact that the NLO corrections are large, as can be seen by comparing them with the LO result.  However, we see that the approximate NLO result shows a smaller scale uncertainty than the NLO result with the contribution of the $qg$ channel. We conclude that the soft emission corrections provide the bulk of the NLO corrections for this choice of the factorization scale. This motivates us to study the effect of the resummation of these corrections, keeping in mind that by matching the resummed results to NLO calculations we consider both power corrections and the contribution of the $qg$ channel to that order.

The NLO+NLL and NLO+NNLL cross sections, shown in the sixth and seventh line of Table~\ref{tab:CSHp13hM} are main results of this paper. By looking at the NLO, NLO+NLL, NLO+NNLL results we see that the cross section is progressively increased, but the central value of each prediction falls in the scale uncertainty band of the predictions of lower accuracy.

One can then look at the NNLO expansions of the NNLL resummation formula, which are shown in the last two lines of Table~\ref{tab:CSHp13hM}. By comparing these results to the NLO+NNLL cross section, one sees that the effect of the resummation corrections beyond NNLO are relatively small. As it was observed in the case of the $t \bar{t} H$ and $t \bar{t} W$ processes in \cite{Broggio:2015lya,Broggio:2016zgg,Broggio:2016lfj}, the scale uncertainty affecting the nNLO result is very small compared to the NLO+NNLL scale uncertainty, and most likely underestimates the residual perturbative uncertainty at NNLO.

\begin{table}[t]
	\begin{center}
		\def\arraystretch{1.3}
		\begin{tabular}{|l|l|c|}
			\hline  $\sqrt{s}$ and pert. order & process & $\sigma$ [fb] \\
			\hline
			\hline  $8$~ TeV NLO & $t \bar{t} W^+$  & $136.7^{+15.6} _{-15.2}$ \\
			\hline  $8$~ TeV NLO & $t \bar{t} W^-$ & $60.5^{+7.1}_{-6.8}$ \\
			\hline  $8$~ TeV NLO & $t \bar{t} Z$ & $189.8^{+24.5}_{-24.8}$ \\
			\hline  $8$~ TeV NLO+NNLL & $t \bar{t} W^+$ & $130.7^{+6.9} _{-4.9}$ \\
			\hline  $8$~ TeV NLO+NNLL & $t \bar{t} W^-$ & $59.1^{+3.1}_{-2.2}$ \\
			\hline  $8$~ TeV NLO+NNLL & $t \bar{t} Z$ & $203.9^{+13.5}_{-15.8}$ \\
			\hline 
			\hline  $13$~ TeV NLO & $t \bar{t} W^+$ & $356.3^{+43.7}_{-39.5} $ \\
			\hline  $13$~ TeV NLO & $t \bar{t} W^-$ & $182.2^{+23.1}_{-20.4} $  \\
			\hline  $13$~ TeV NLO & $t \bar{t} Z$ & $728.3^{+93.8}_{-90.3}$  \\
			\hline  $13$~ TeV NLO+NNLL & $t \bar{t} W^+$ & $341.0^{+23.1}_{-13.6} $ \\
			\hline  $13$~ TeV NLO+NNLL & $t \bar{t} W^-$ & $177.1^{+12.0}_{-6.9} $ \\
			\hline  $13$~ TeV NLO+NNLL &$ t \bar{t} Z$ & $777.8^{+61.3}_{-65.2} $ \\
			\hline 
		\end{tabular} 
		\caption{Total cross section for $t \bar{t} Z$ and $t \bar{t} W$ production at the
			LHC with $\sqrt{s} = 8$ and $13$~TeV and MMHT 2014 PDFs. The default value
			of the factorization scale is $\mu_{f,0}=M/2$, and
			the uncertainties are estimated through
			variations of this scale (and of the resummation scales
			$\mu_s$ and $\mu_h$ when applicable).
			\label{tab:ttZttW}}
	\end{center}
\end{table}

Experimental collaborations reported measurements of the $t \bar{t} Z$ total cross section in combination with measurements of the  $t \bar{t} W$  cross section \cite{Khachatryan:2015sha,ATLAS13,CMS:2016dui,Aad:2015eua}. We conclude this section by comparing our predictions for $t \bar{t} W$ and $t \bar{t} Z$ with experimental data. The  $t \bar{t} W$ production cross section was evaluated by running the code developed in \cite{Broggio:2016zgg} with the scale choices and input parameters employed in the present work for $t \bar{t} Z$ production and described above. The results for $t \bar{t} Z$ and $t \bar{t} W$ production cross section at $8$ and $13$~TeV are summarized in Table~\ref{tab:ttZttW}. 
\begin{figure}[tp]
	\begin{center}
		\begin{tabular}{cc}
			\includegraphics[width=7.2cm]{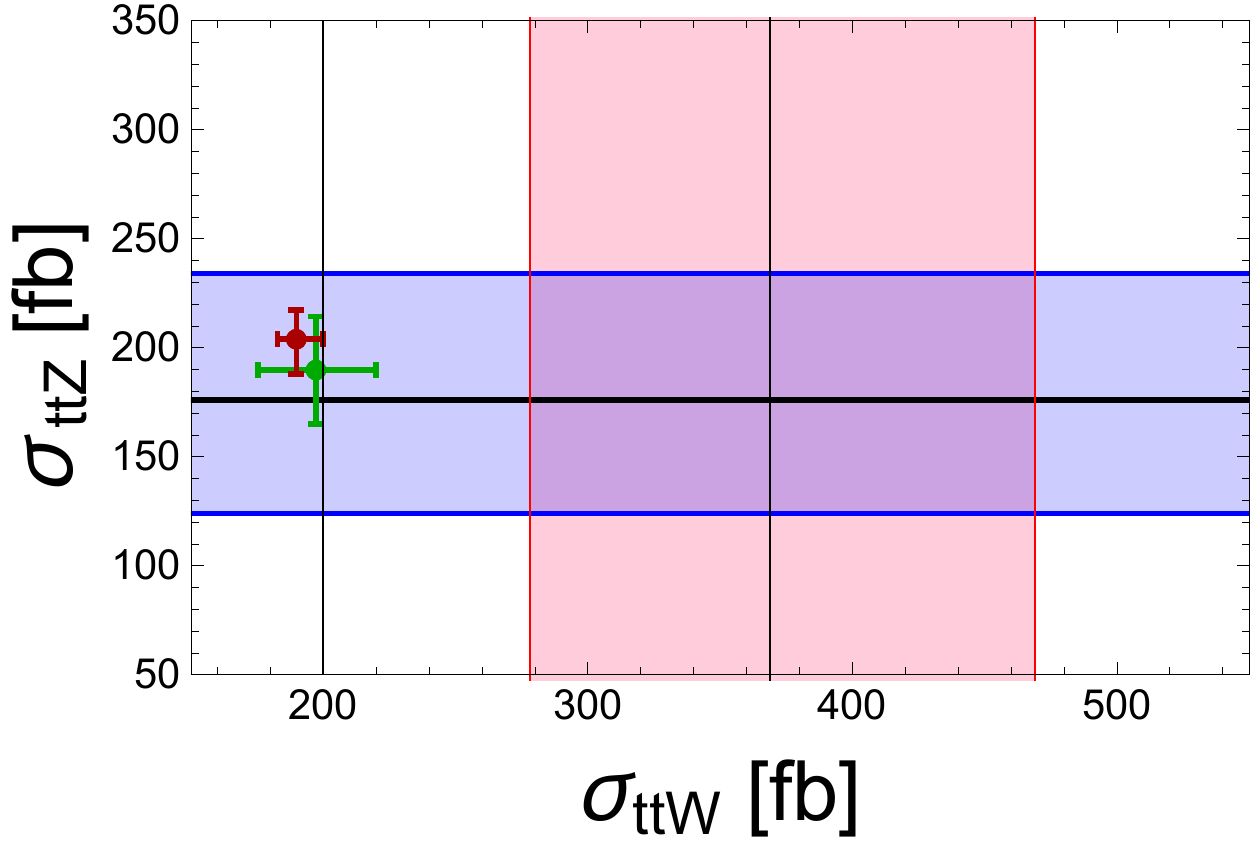} & \includegraphics[width=7.2cm]{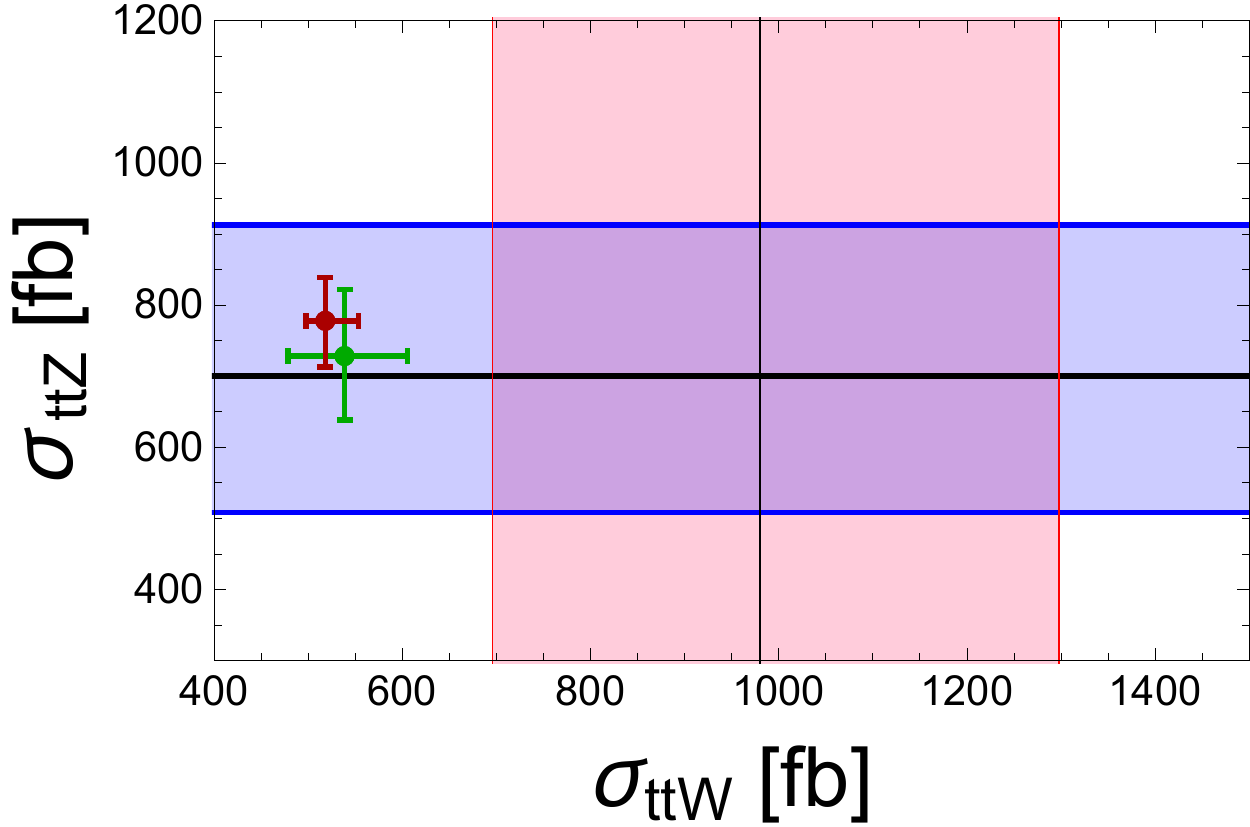} \\
		\end{tabular}
	\end{center}
	\caption{Total cross section at NLO (Green) and NLO+NNLL (Red) compared to the ATLAS measurements at $8$~TeV \cite{Aad:2015eua} (left panel) and CMS measurement at $13$~TeV \cite{CMS:2016dui} (right panel). \label{fig:totCS8and13}}
	
\end{figure}
In Figure~\ref{fig:totCS8and13} we follow the structure of Figure~12 in~\cite{CMS:2016dui} in order to   compare graphically  calculations with the corresponding experimental measurements. The experimental measurements at $8$~TeV are taken from \cite{Aad:2015eua}, while the experimental measurements at $13$~TeV are taken from \cite{CMS:2016dui}. The green dots and cross-shaped ``error bars'' correspond to NLO calculations carried out with $\mu_{f,0} = M/2$ and their scale uncertainty. The red dots and crosses correspond instead to NLO+NNLL calculations. 
%

It is interesting to observe that, while predictions for the $t \bar{t}Z$ production cross section are in perfect agreement with the measurements at both $8$ and $13$~TeV, the predictions for the $t \bar{t} W$ cross section are slightly smaller than measurements for both collider energies. This observation holds for NLO and NLO+NNLL calculations alike. Of course this discrepancy should be taken with a grain of salt, and requires a more detailed discussion with the experimental collaborations. Moreover, we would like to stress that a fully exhaustive comparison between predictions and measurements should also account for the uncertainty associated to the choice of the PDFs and to the value of $\alpha_s$. These two sources of uncertainty are not reflected in the error bars of Figure~\ref{fig:totCS8and13}.
\subsection{Differential distributions}
\label{sec:diffdist}

\begin{figure}[tp]
	\begin{center}
		\begin{tabular}{cc}
			\includegraphics[width=7.2cm]{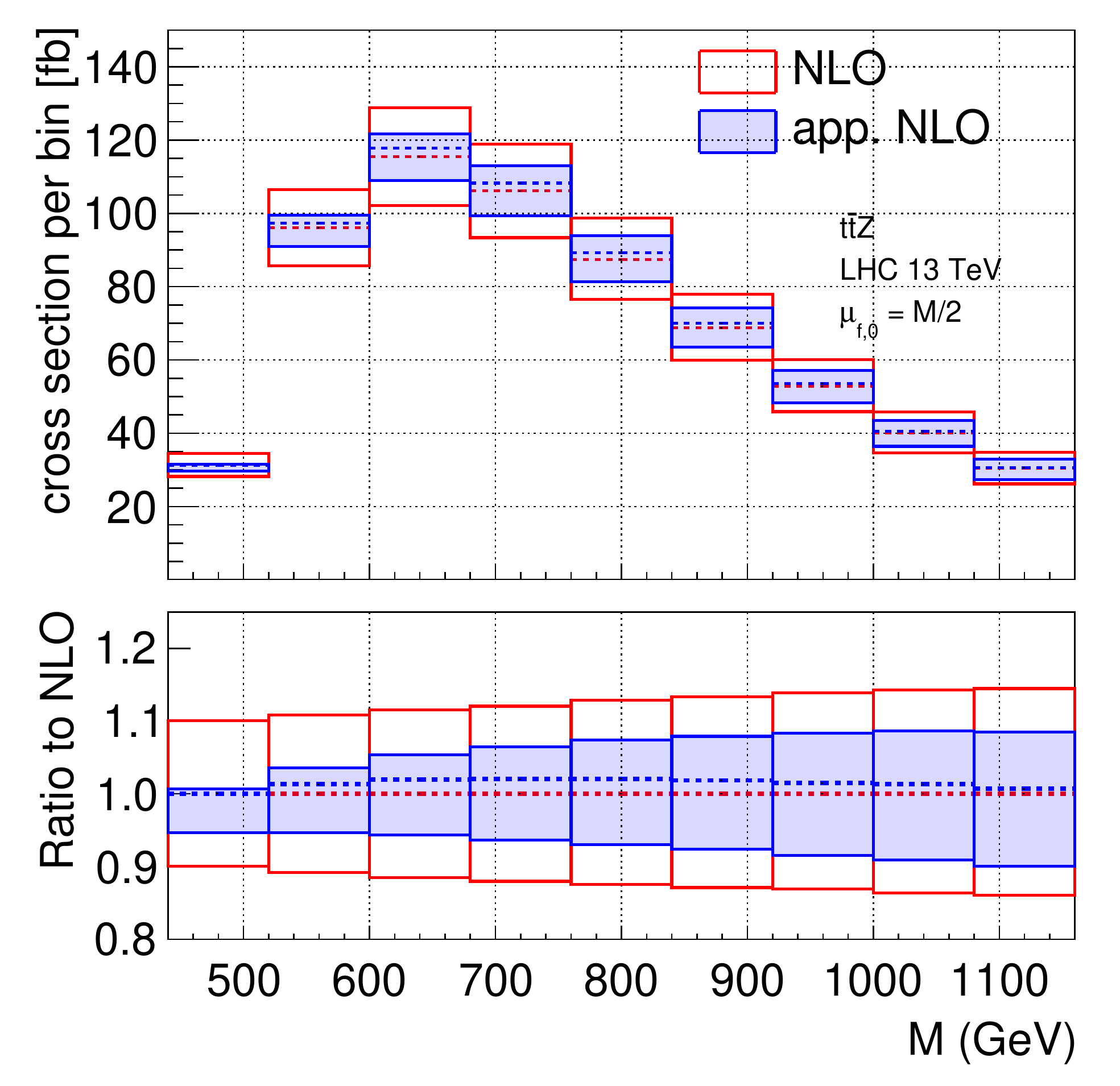} & \includegraphics[width=7.2cm]{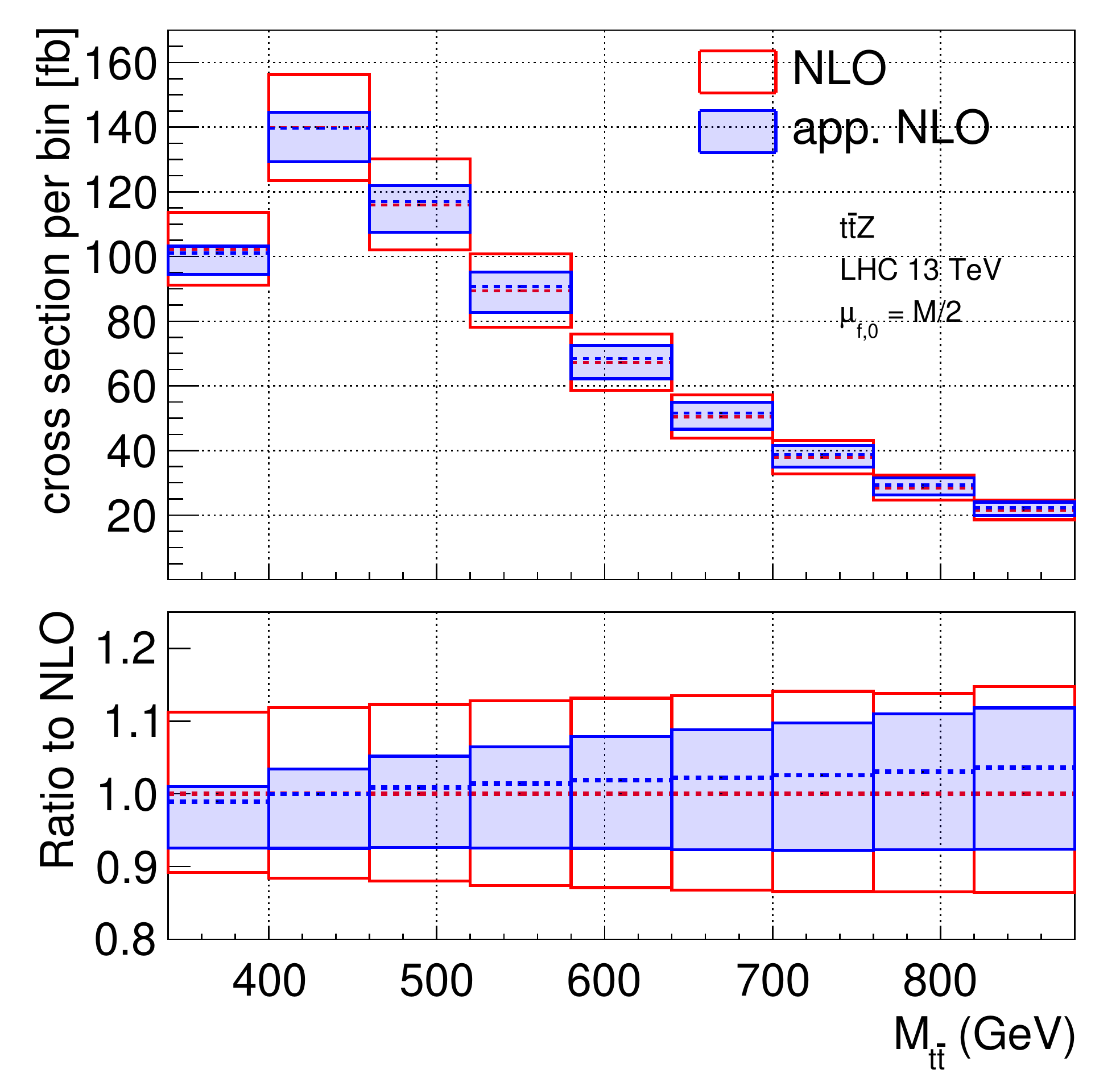} \\
			\includegraphics[width=7.2cm]{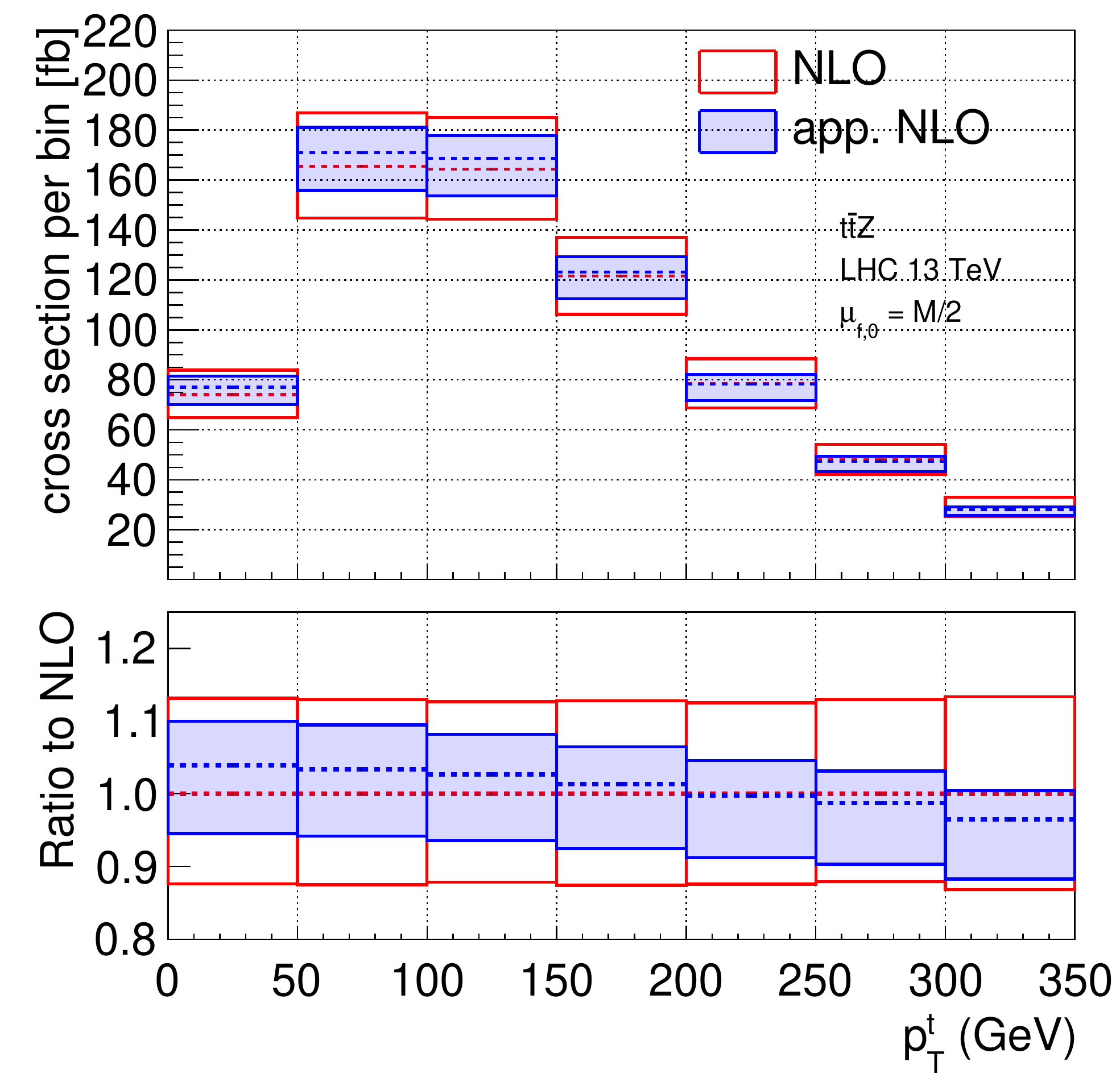} & \includegraphics[width=7.2cm]{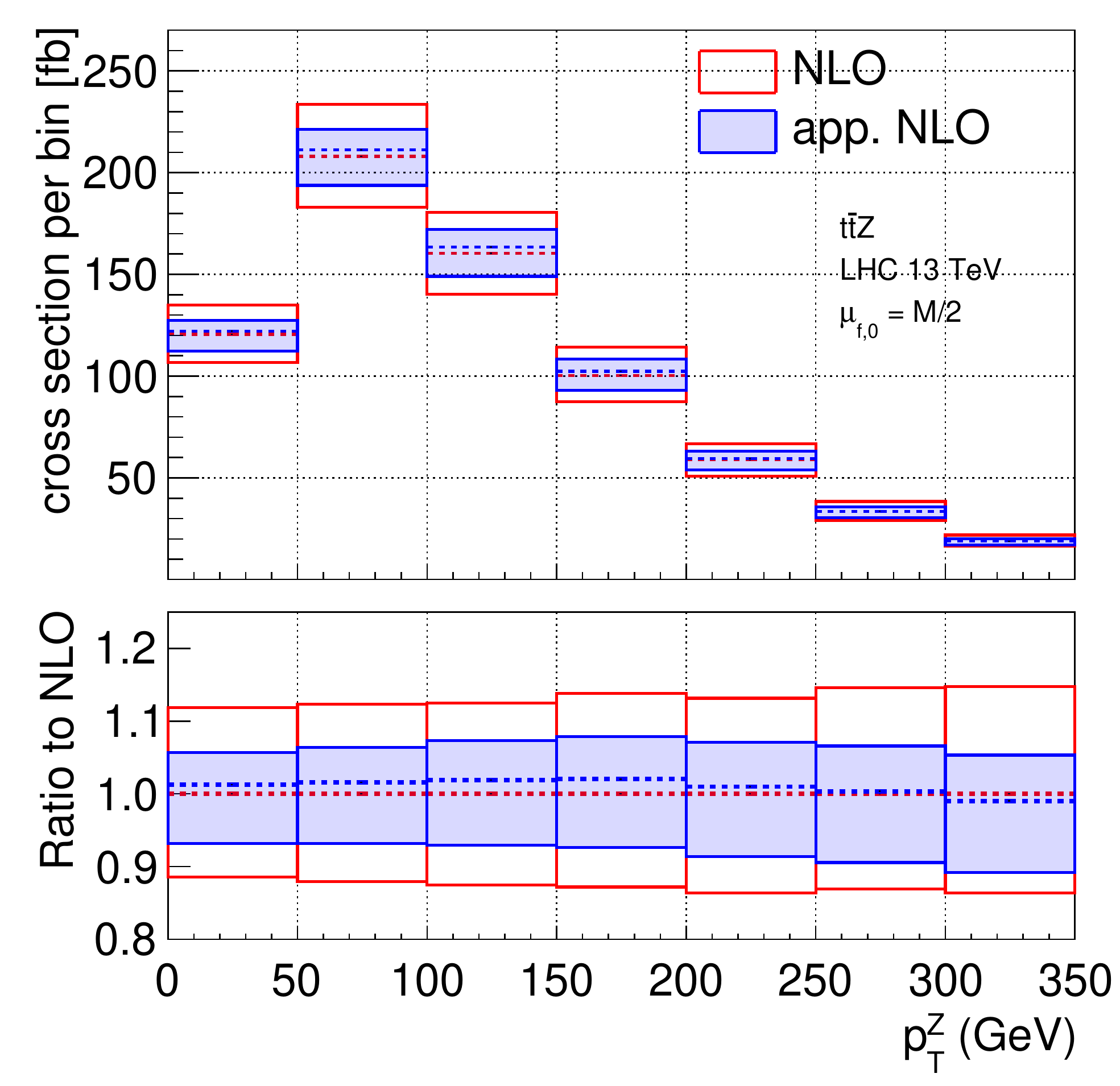} \\
		\end{tabular}
	\end{center}
	\caption{Differential distributions at approximate NLO (blue
		band) compared to the complete NLO (red band). The default
		factorization scale is chosen as $\mu_{f,0}=M/2$, and the
		uncertainty bands are generated through scale variations as
		explained in the text. \label{fig:nLOvsNLOhalfM}}
	
\end{figure}

\begin{figure}[tp]
	\begin{center}
		\begin{tabular}{cc}
			\includegraphics[width=7.2cm]{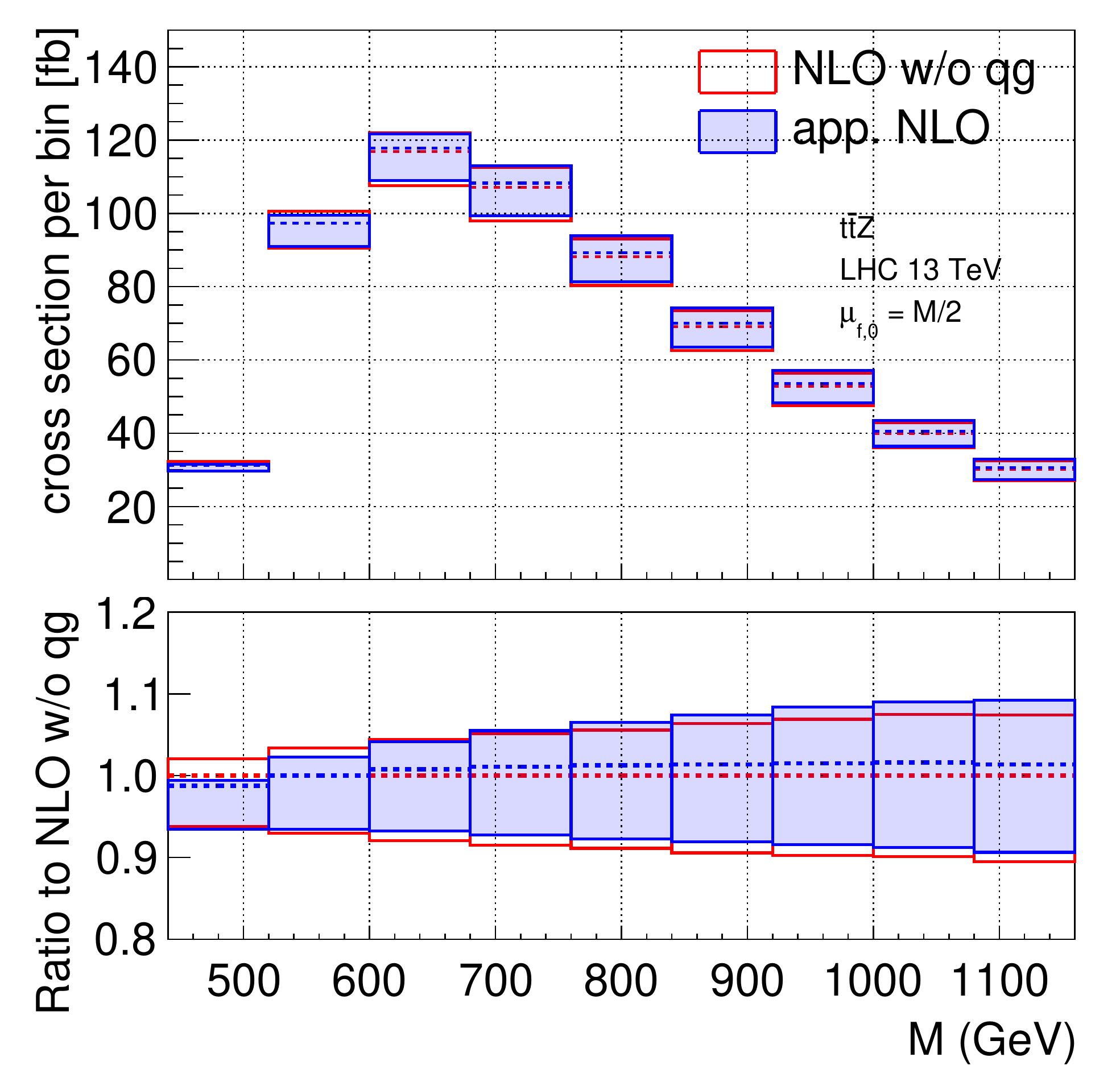} & \includegraphics[width=7.2cm]{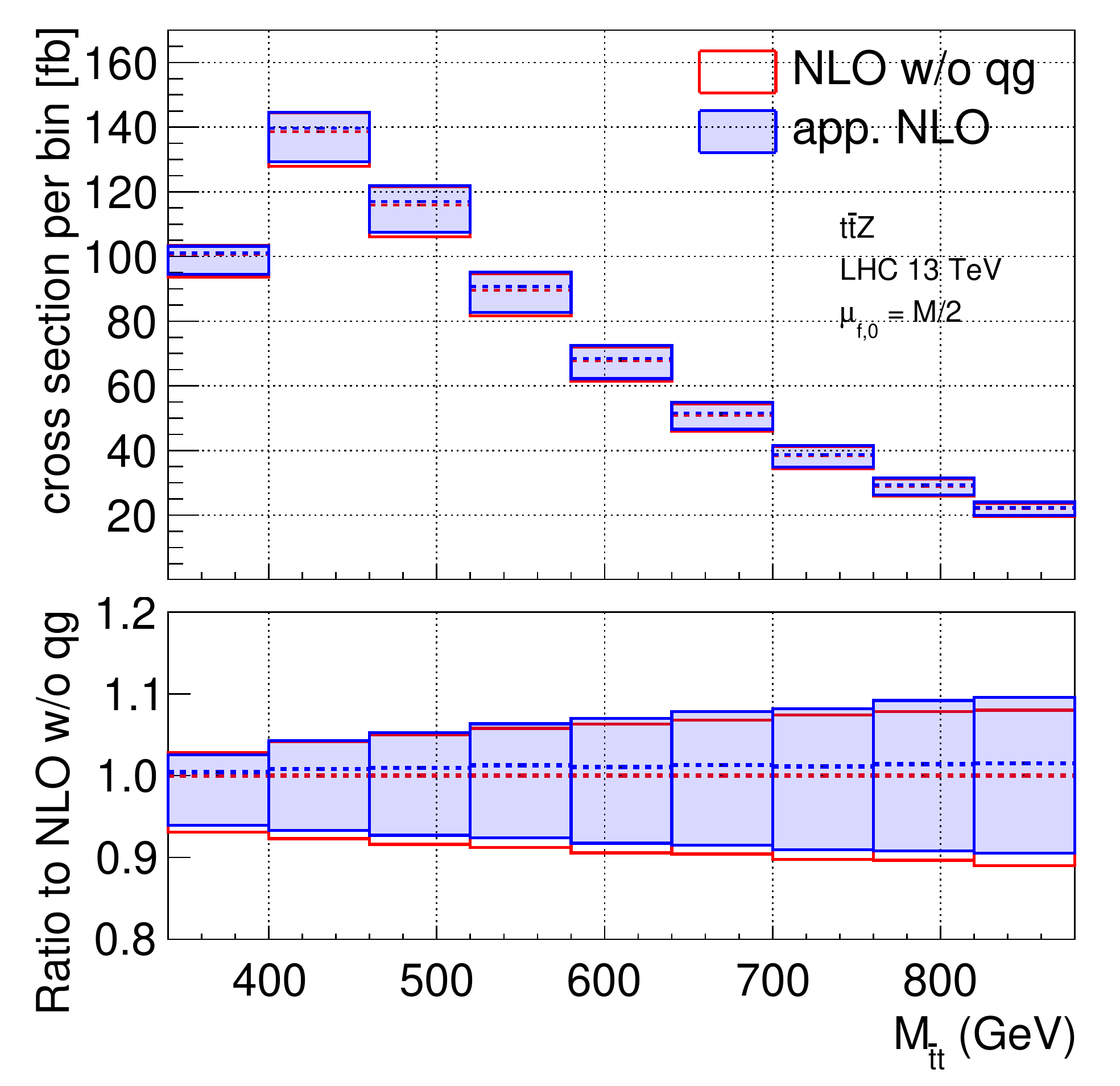} \\
			\includegraphics[width=7.2cm]{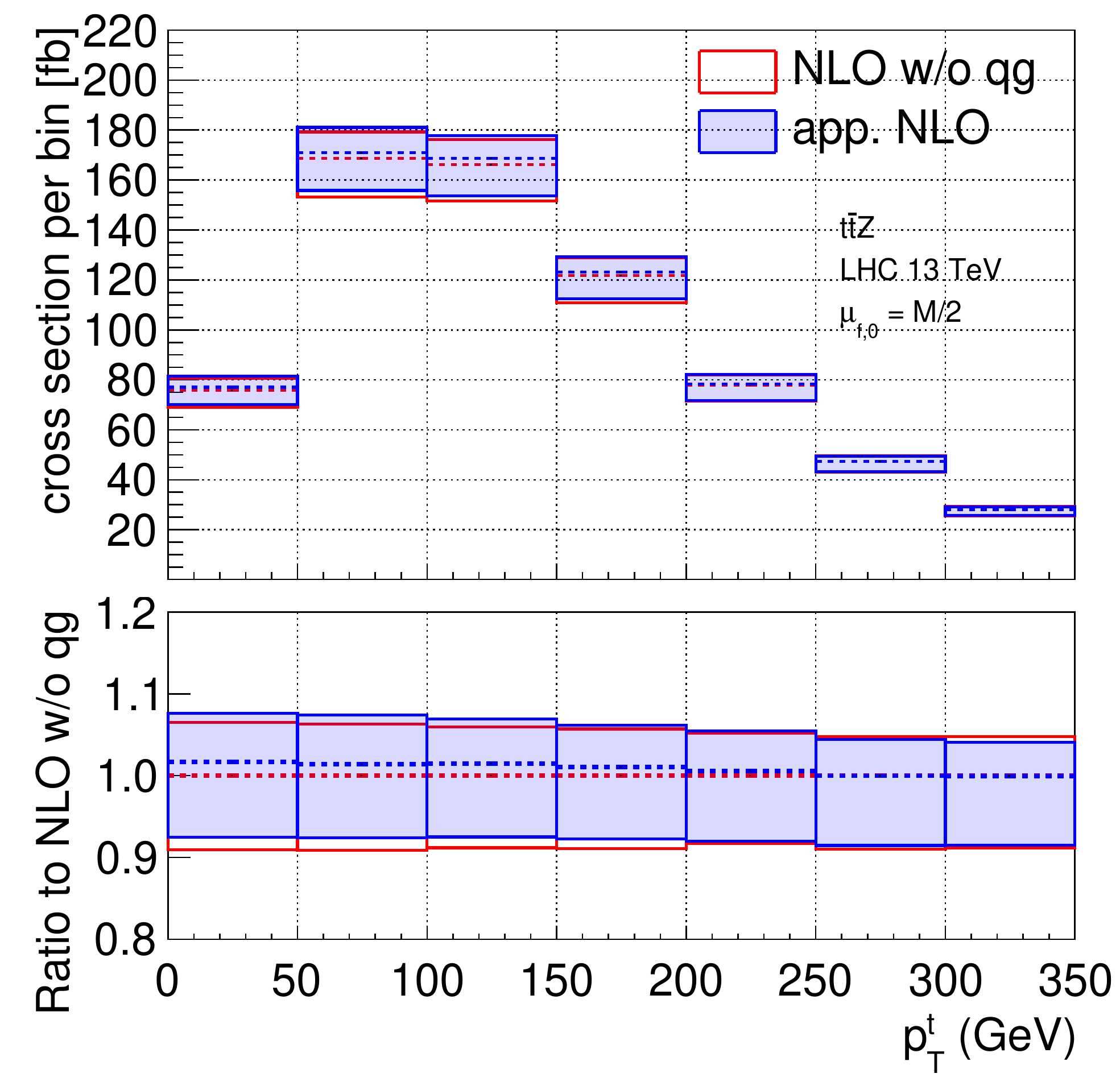} & \includegraphics[width=7.2cm]{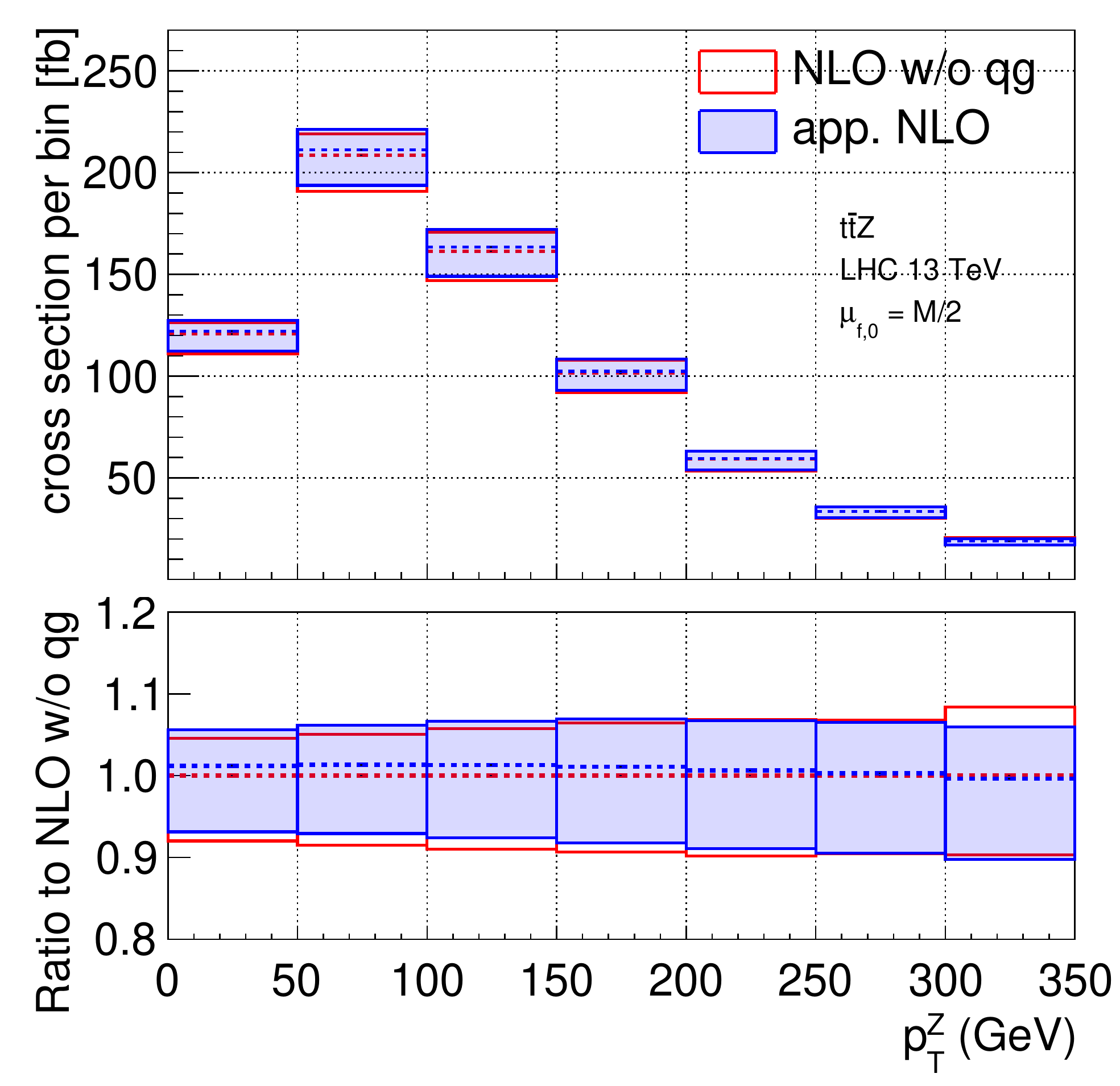} \\
		\end{tabular}
	\end{center}
	\caption{Differential distributions at approximate NLO (blue band) compared to the NLO distributions without the quark-gluon channel contribution (red band). All settings are as in Figure~\ref{fig:nLOvsNLOhalfM}.
		\label{fig:nLOvsNLOnoqghaflM}
	}
	
\end{figure}

In this section we obtain predictions for four differential distributions which depend on the momenta of the final state massive particles. The distributions are \emph{i)} the distribution differential with respect to the $t \bar{t} Z$ invariant mass, $M$, \emph{ii)} the distribution differential with respect to the $t \bar{t}$ invariant mass, $M_{t \bar{t}}$, \emph{iii)}  the distribution differential with respect to the transverse momentum of the top quark, $p^t_T$, and  \emph{iv)} the distribution differential with respect to the transverse momentum of the $Z$ boson, $p^Z_T$. 

Figure~\ref{fig:nLOvsNLOhalfM} compares the approximate NLO calculations, carried out with our in-house code, with the complete NLO calculations, carried out with \mgamc. We see that the approximate NLO calculations reproduce well the full NLO calculations. The lower part of each panel shows the ratio between the approximate NLO or complete NLO calculations and  the central value of the NLO calculation. One can see that the approximate NLO scale uncertainty band is included in the NLO scale uncertainty band. Figure~\ref{fig:nLOvsNLOnoqghaflM} repeats the same analysis but it compares approximate NLO calculations to NLO calculations without the quark-gluon channel contribution. As expected approximate NLO distributions and NLO distributions without the $qg$ channel have the same shape and scale uncertainty bands of similar size. These two figures show that, for this choice of the factorization scale at least, soft emission corrections provide the bulk of the NLO corrections.

\begin{figure}[tp]
	\begin{center}
		\begin{tabular}{cc}
			\includegraphics[width=7cm]{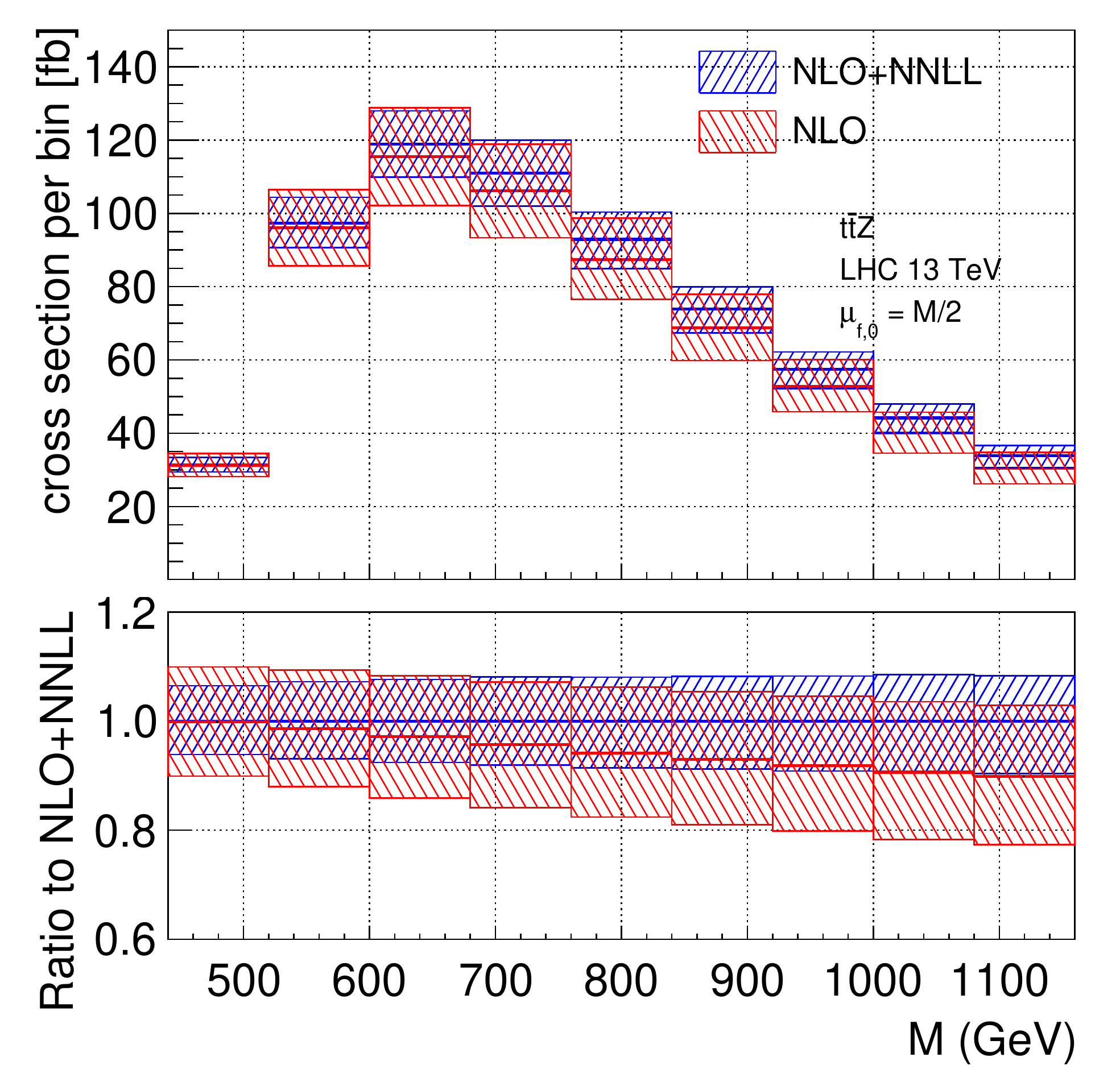} & \includegraphics[width=7cm]{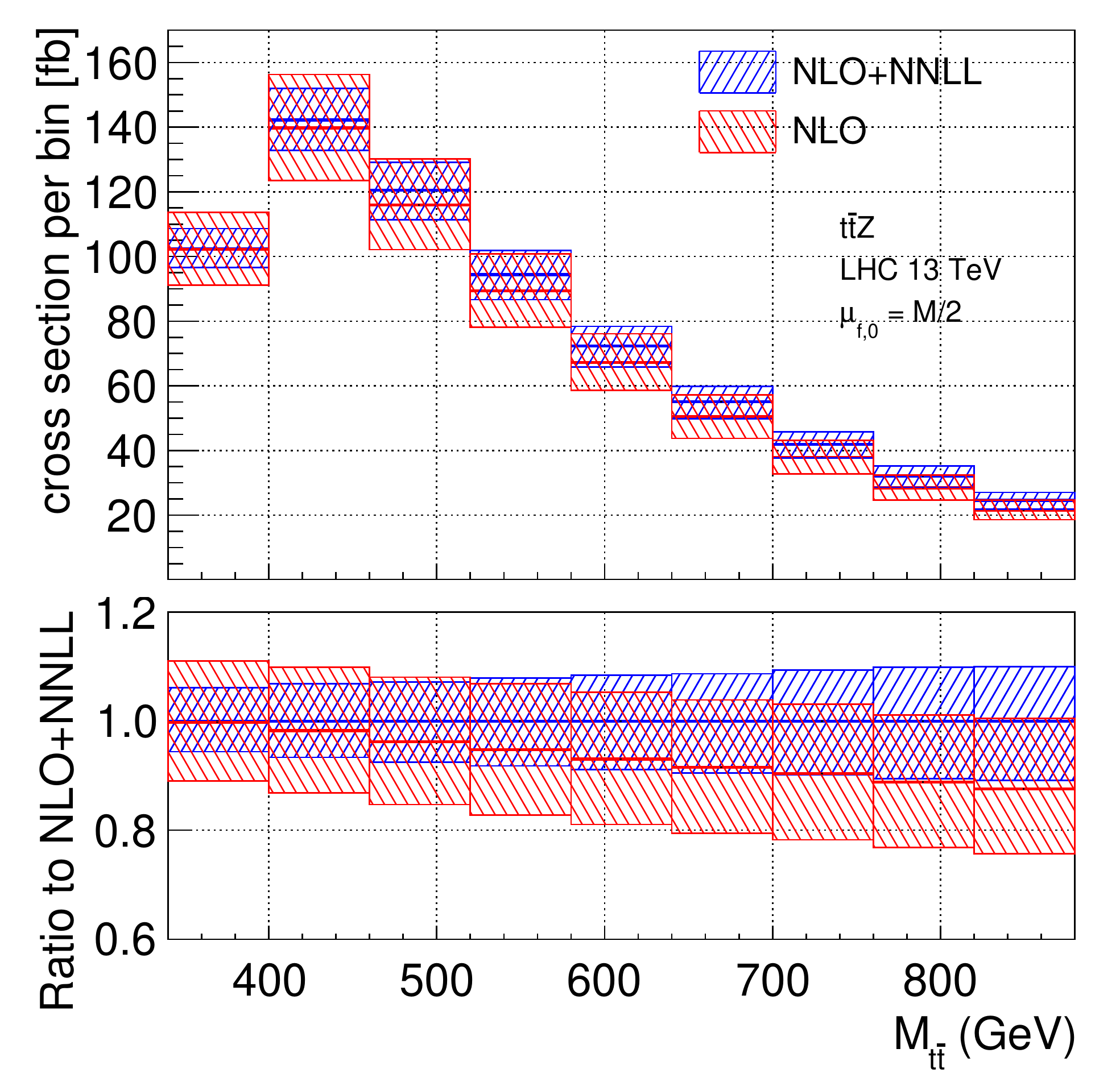} \\
			\includegraphics[width=7cm]{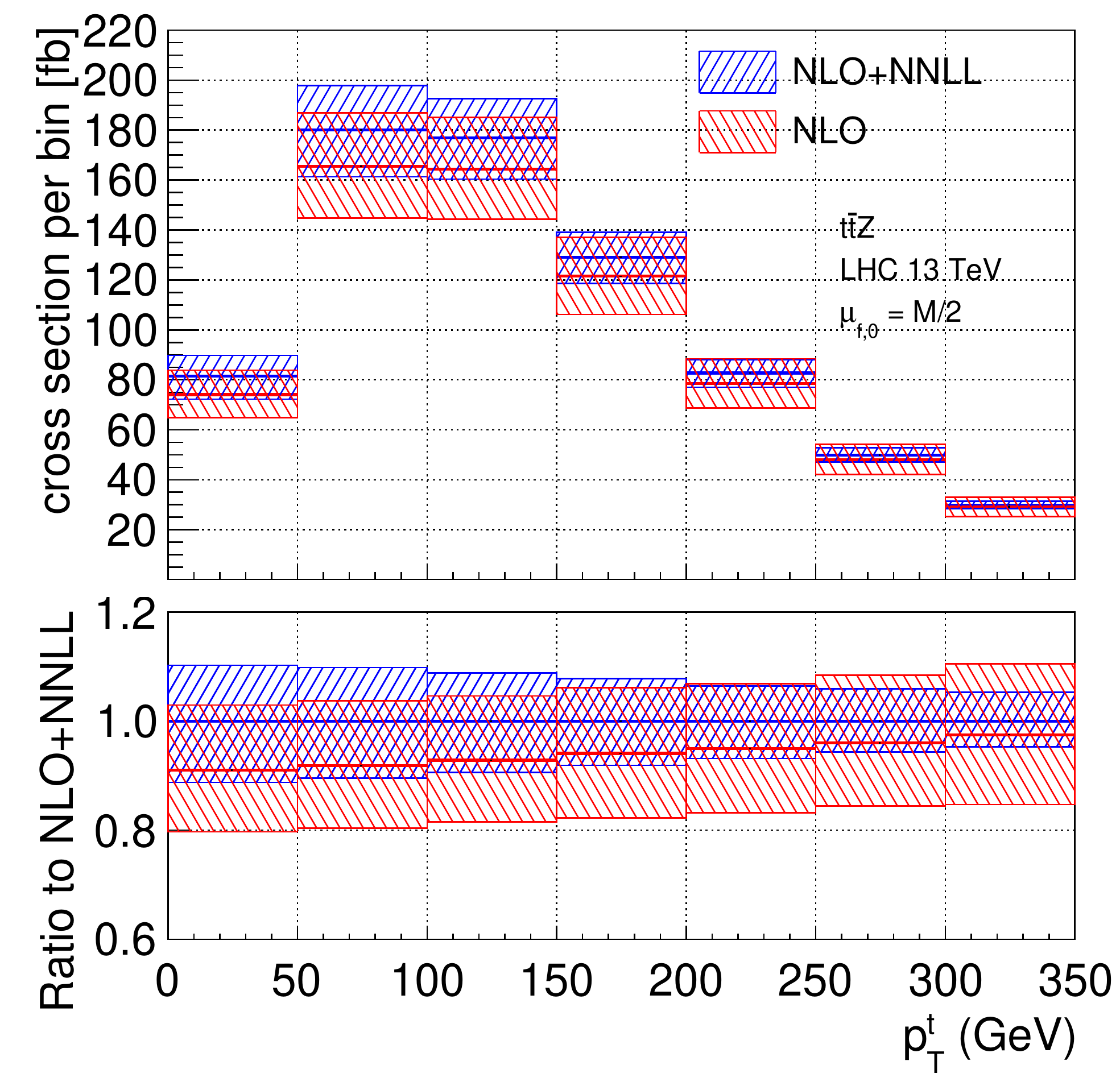} & \includegraphics[width=7cm]{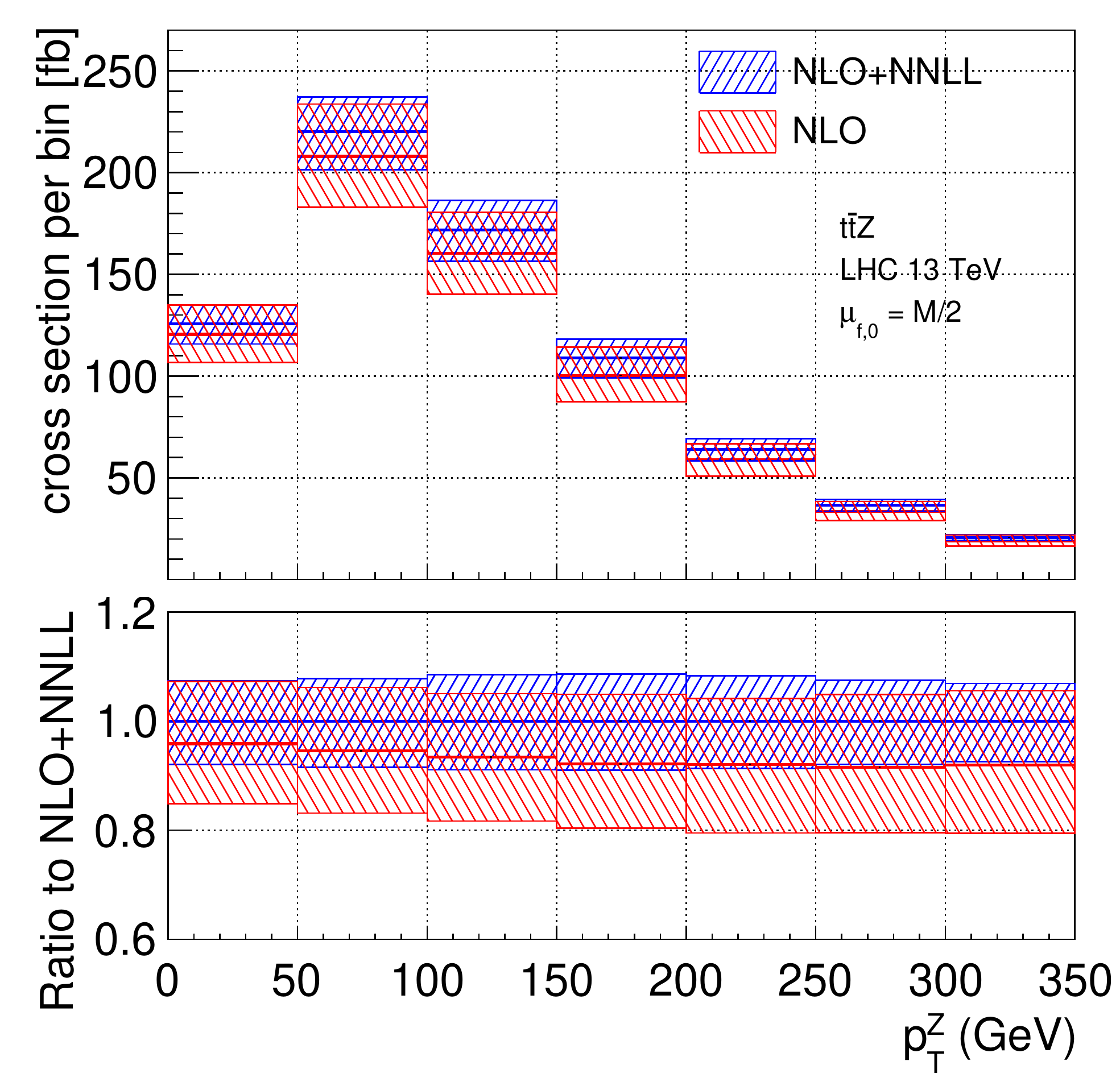} \\
		\end{tabular}
	\end{center}
	\caption{Differential distributions with $\mu_{f,0}=M/2$ at NLO+NNLL (blue band) compared to the NLO calculation (red band). The uncertainty bands
		are generated through scale variations of $\mu_f$, $\mu_s$ and $\mu_h$ as 
		explained in the text.
		\label{fig:NLOvsNNLLhalfM}
	}
\end{figure}

\begin{figure}[tp]
	\begin{center}
		\begin{tabular}{cc}
			\includegraphics[width=7cm]{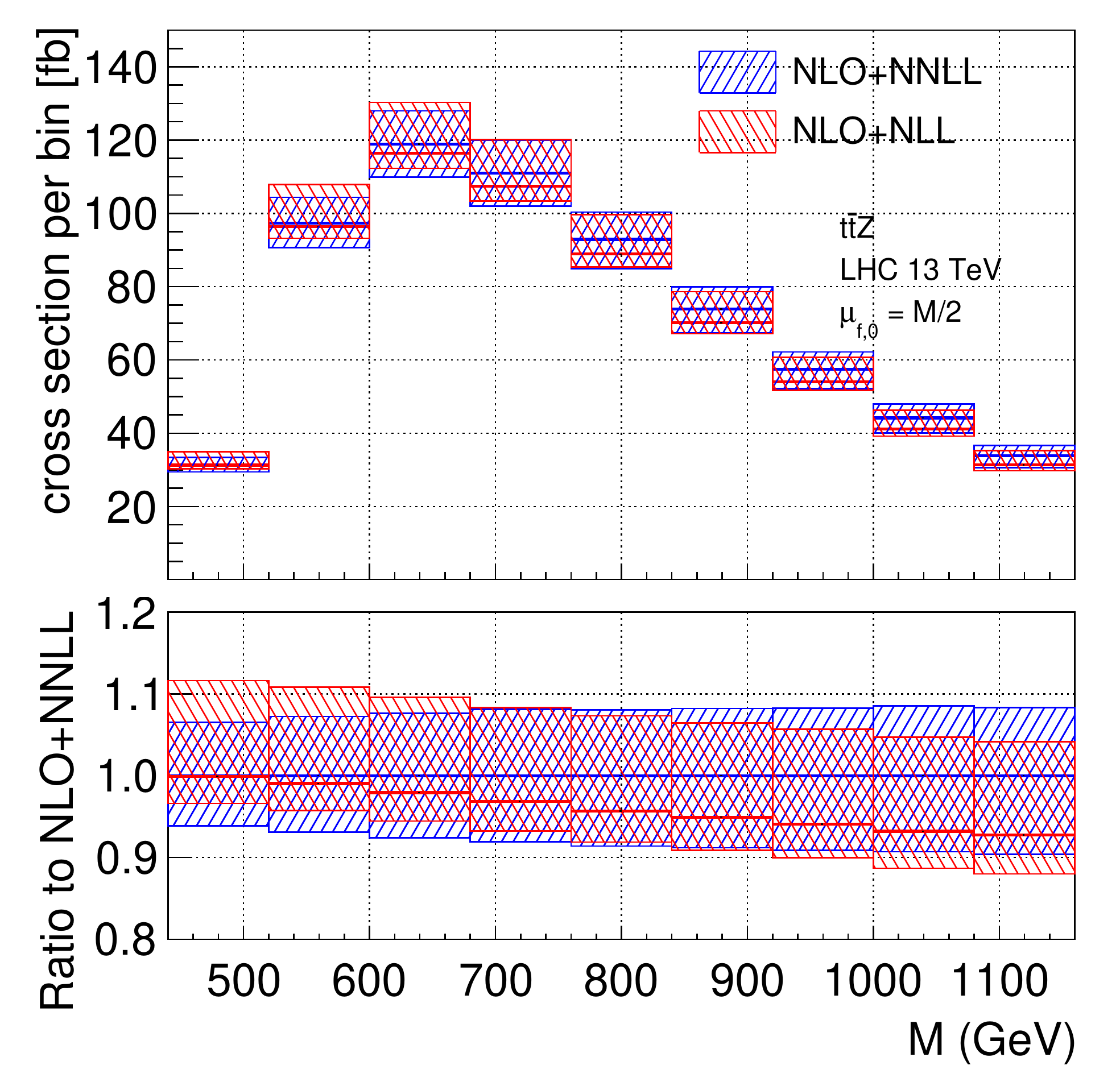} & \includegraphics[width=7cm]{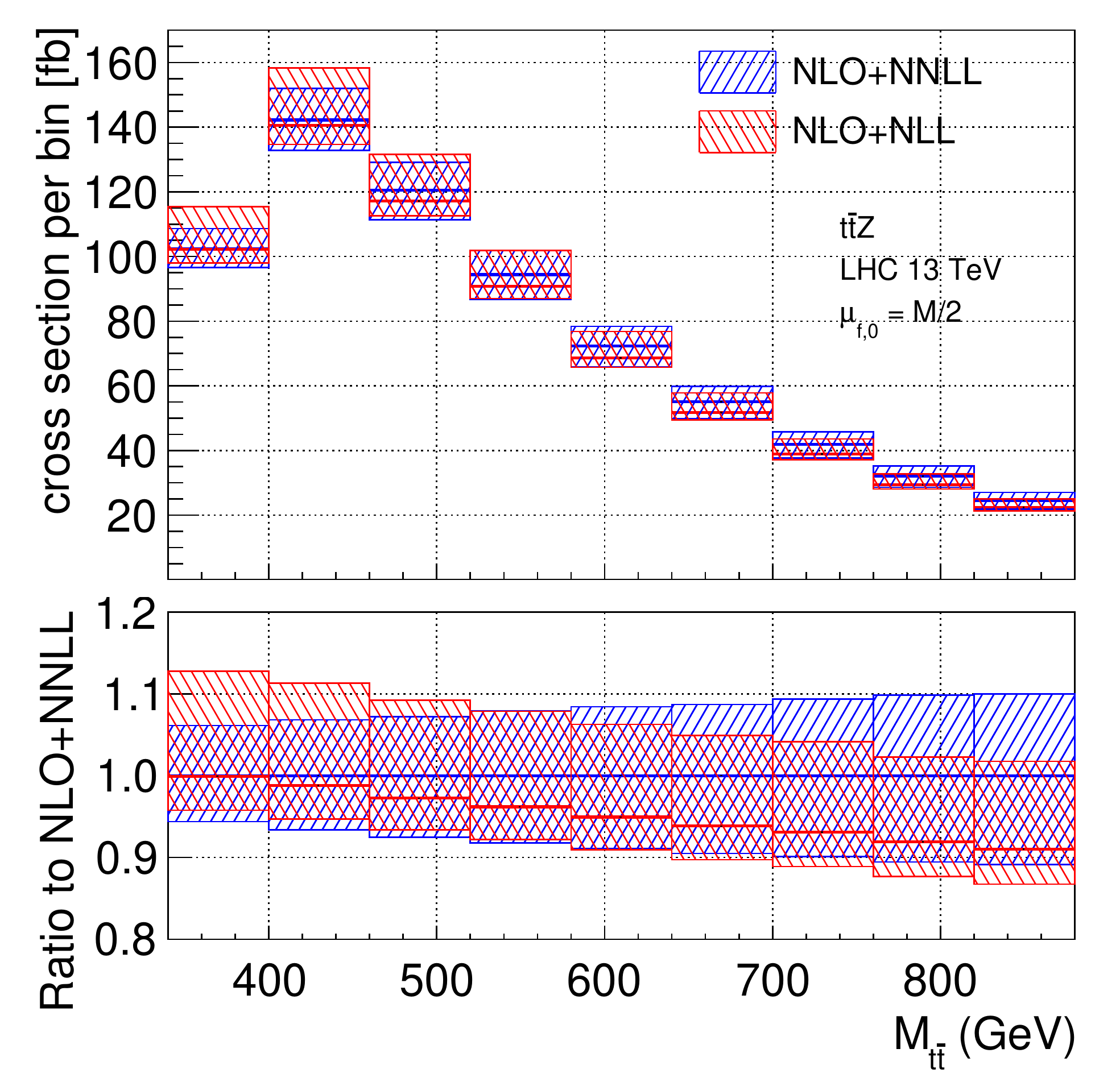} \\
			\includegraphics[width=7cm]{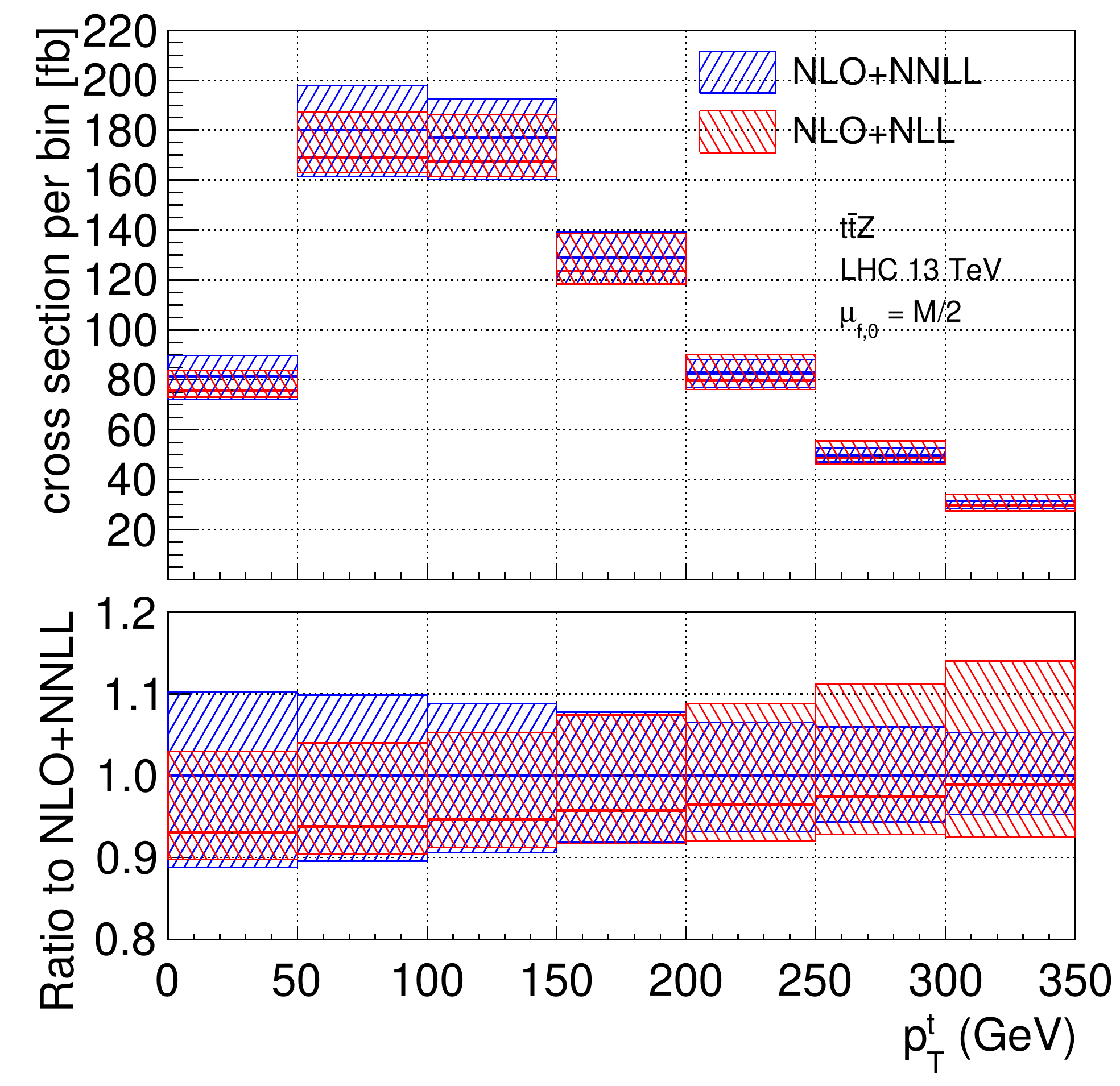} & \includegraphics[width=7cm]{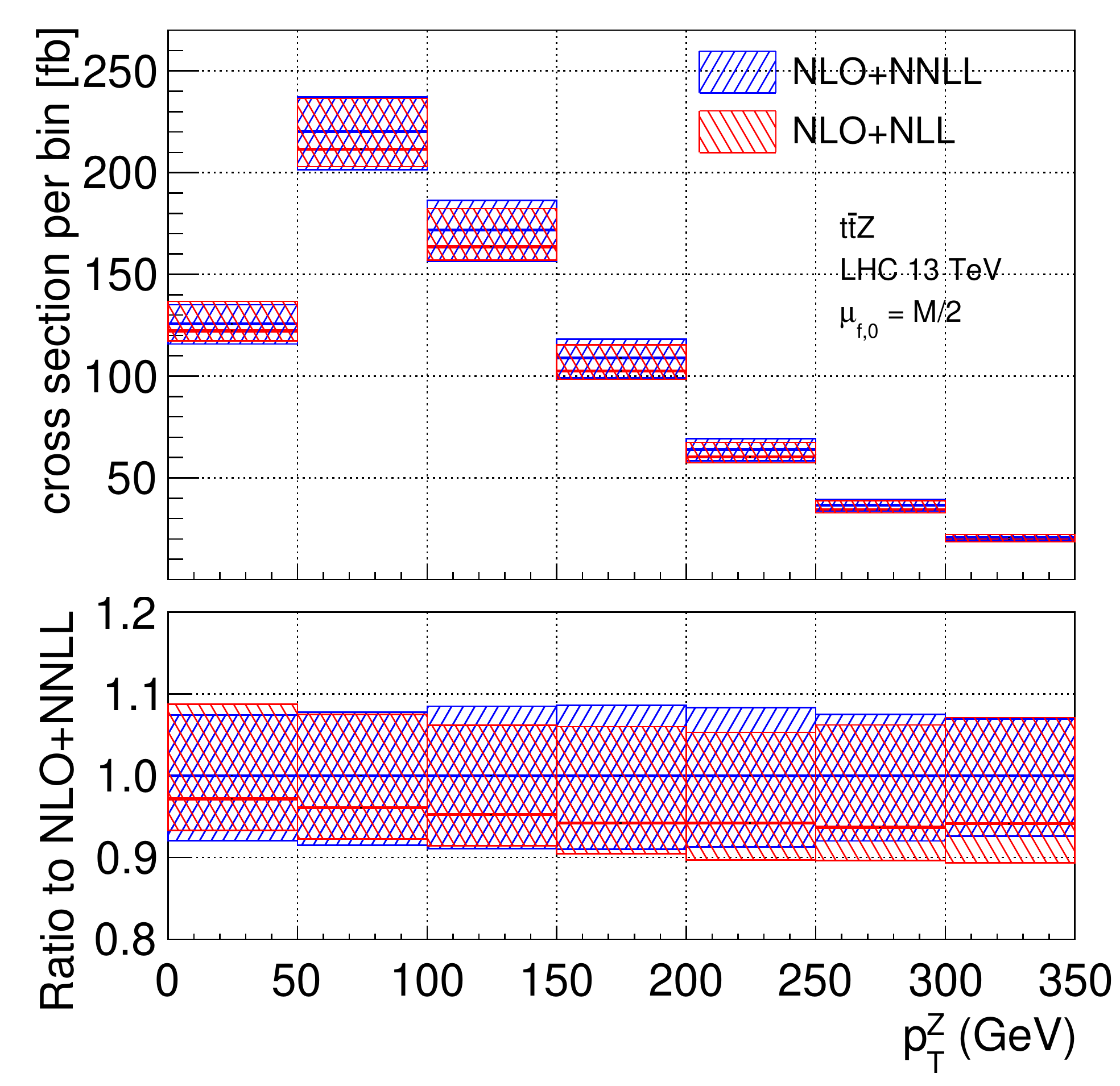} \\
		\end{tabular}
	\end{center}
	\caption{Differential distributions  with $\mu_{f,0}=M/2$ at NLO+NNLL (blue band) compared to the corresponding NLO+NLL calculation (red band).   The uncertainty bands
		are generated through scale variations.
		\label{fig:NLLvsNNLLhalfM}
	}
\end{figure}

\begin{figure}[tp]
	\begin{center}
		\begin{tabular}{cc}
			\includegraphics[width=7cm]{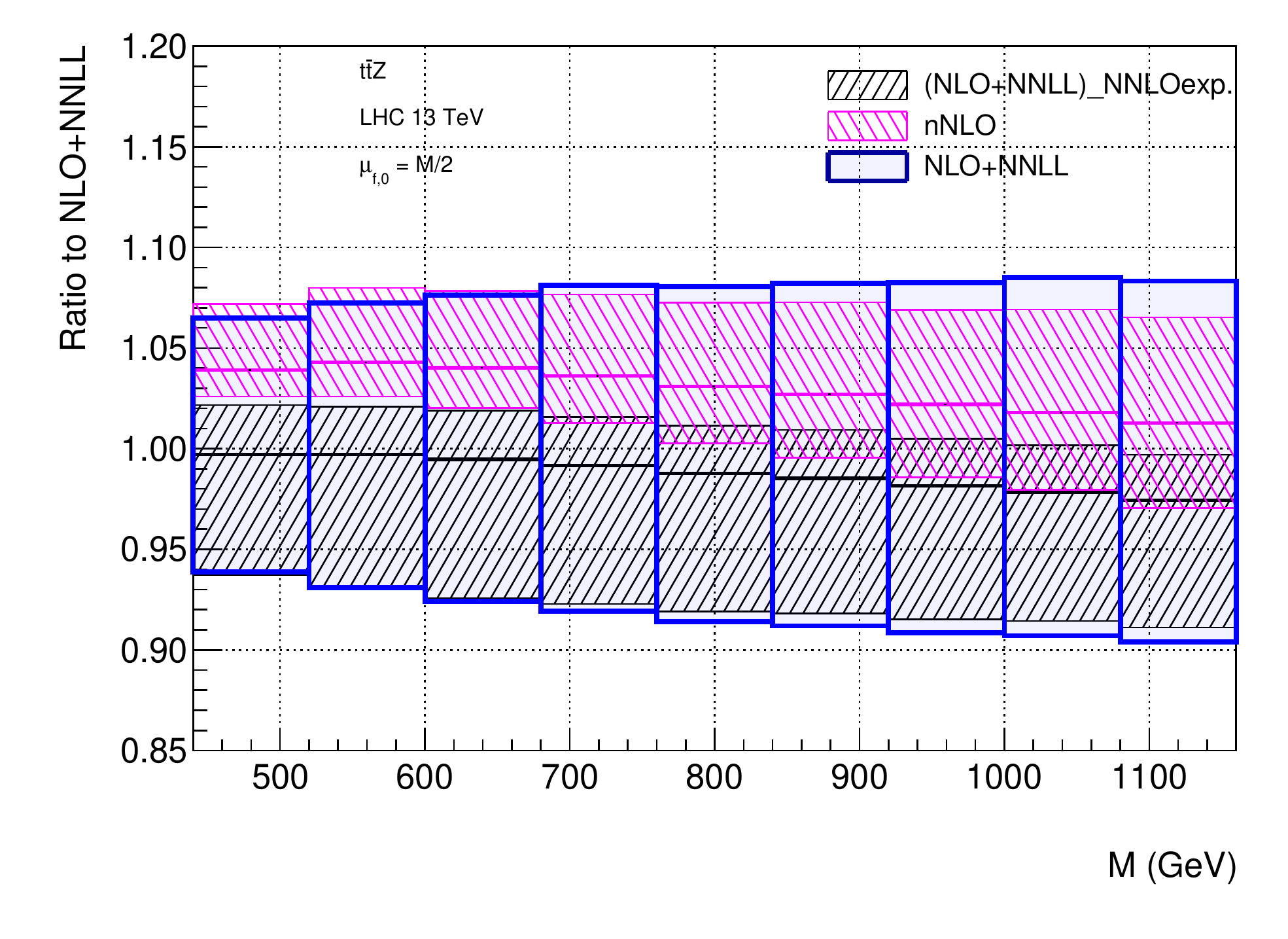} & \includegraphics[width=7cm]{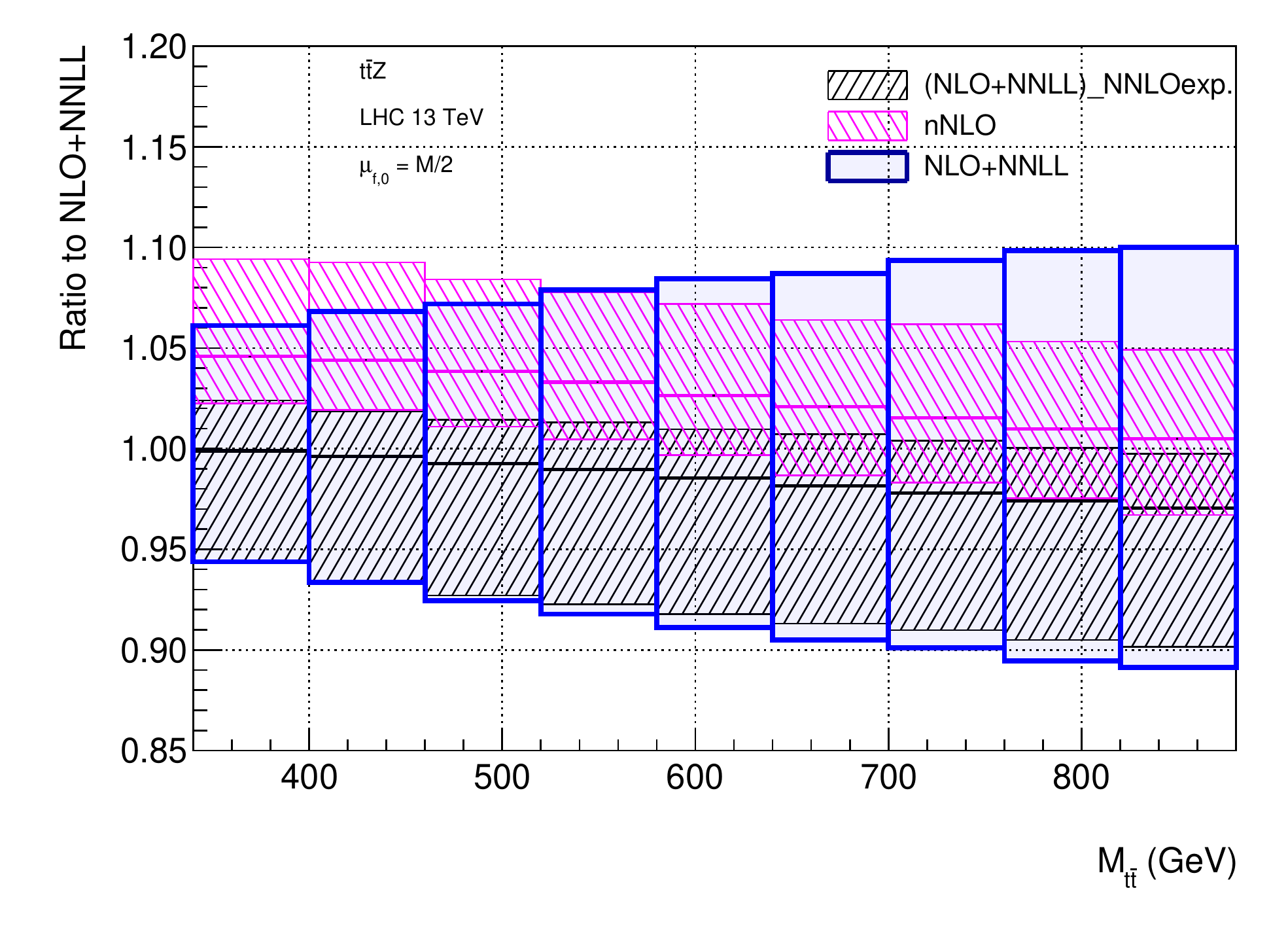} \\
			\includegraphics[width=7cm]{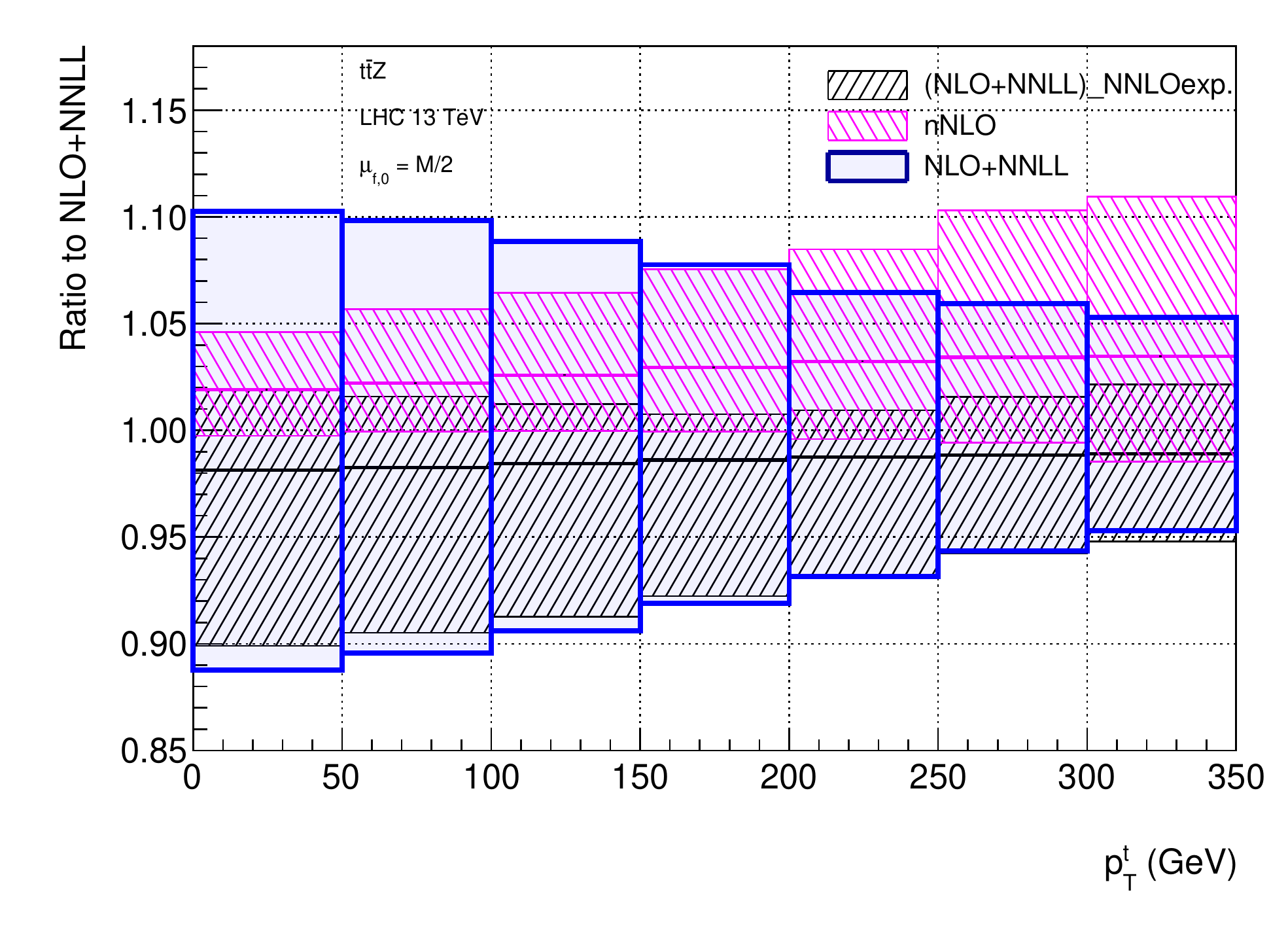} & \includegraphics[width=7cm]{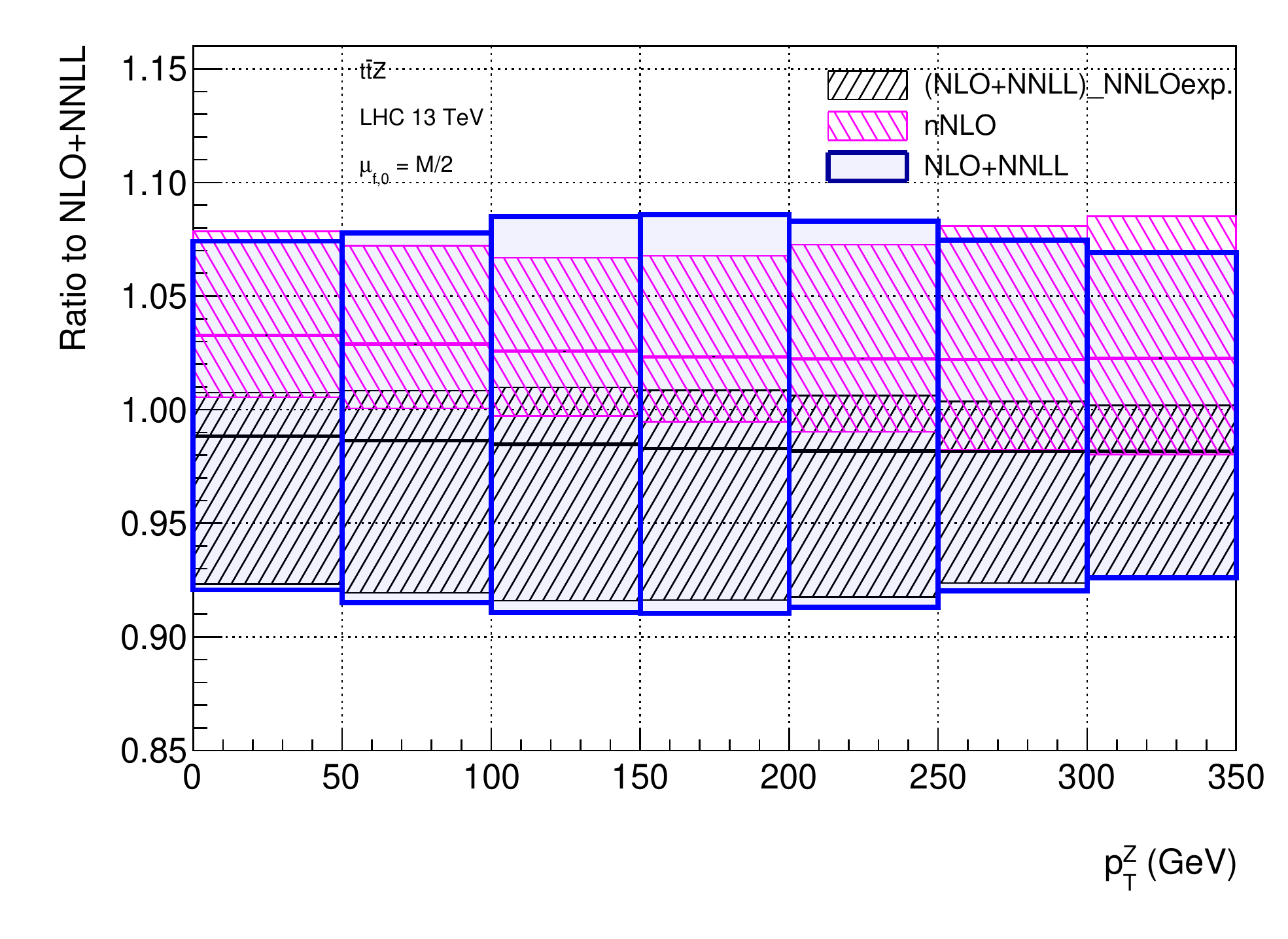} \\
		\end{tabular}
	\end{center}
	\caption{Differential distributions ratios for $\mu_{f,0}=M/2$, where 
		the uncertainties are generated through scale variations.
		\label{fig:NNLOratshalfM}
	}
\end{figure}


Figure~\ref{fig:NLOvsNNLLhalfM} provides the main result of this section. This figure compares NLO calculations to the distributions evaluated to NLO+NNLL accuracy. Roughly, one can say that the NLO+NNLL results fall in the upper part of the NLO scale uncertainty interval in each bin. The central value of the NLO+NNLL calculations is slightly larger than the central value of the NLO calculations in all bins shown. The scale uncertainty affecting the NLO+NNLL accuracy calculation, which is obtained by 
varying $\mu_s$, $\mu_f$, and $\mu_h$ as described above, is smaller than the NLO scale uncertainty band obtained 
by varying $\mu_f$.

Results at NLO+NLL and NLO+NNLL accuracy are compared in Figure~\ref{fig:NLLvsNNLLhalfM}. The main effect of the NNLL correction with respect to the NLL ones is an increase of the central value of the bins in the tail of the $M$ and $M_{t \bar{t}}$ distributions. The scale uncertainty bands turn out to be of similar size at NLO+NLL and NLO+NNLL in almost all bins shown.

Figure~\ref{fig:NNLOratshalfM} shows the ratio of distributions at various level of precision to the central value of the NLO+NNLL calculation in each bin.
In particular, the blue band refers to NLO+NNLL distributions, the dashed red band to nNLO distributions
and the dashed black band to distributions obtained from the NNLO expansion of the NLO+NNLL resummation. The NLO+NNLL expanded distributions  differ from the NLO+NNLL distributions by NNLL resummation
effects of order N$^3$LO and higher. These corrections can be as large as 5 to 10 \% in all bins shown, and are particularly relevant at higher values of $\mu_f$.  
The difference between the nNLO and the NLO+NNLL expanded to NNLO results is due to constant NNLO terms, which are formally of order N$^3$LL. Both the NNLO expansion of the NLO+NNLL calculation and the nNLO calculation underestimate the scale uncertainty which one finds at NLO+NNLL accuracy, a fact which we already observed by looking at the predictions for the total cross section. The envelope of
the two NNLO approximations (i.e. the black and red bands) spans almost all of the  NLO+NNLL scale uncertainty interval in each bin, with the exception of the tail of the $p^t_T$ distribution, where this envelope includes the NLO+NNLL scale uncertainty. 

\section{Conclusions}
\label{sec:conclusions}

In the present work we carried out the resummation of soft gluon emission corrections to the associated production of a top-antitop quark pair and a $Z$ boson. The resummation was studied in the partonic threshold limit $z \to 1$ and was implemented to NNLL accuracy. Numerical calculations of the total cross section and differential distributions to NNLL accuracy were carried out by means of an in-house partonic Monte Carlo code which we developed for this work. The output of this code was matched with NLO calculations obtained from \mgamc. The final outcome of this work is represented by the NLO+NNLL calculations of the total cross section and differential distributions for the LHC operating at a center-of-mass energy of $13$~TeV presented in the previous section. The code can be easily adapted to carry out phenomenological studies which include cuts on the top, antitop and/or $Z$ boson momenta. 

With the choice of the factorization scale made in this work, we can conclude that the soft emission corrections to $t \bar{t} Z$ production evaluated to NNLL accuracy lead to a moderate increase of the total cross section and differential distributions with respect to NLO calculations of the same observables. The residual perturbative uncertainty at NLO+NNLL accuracy, estimated by varying the soft, hard and factorization scales as explained in the text, is smaller than the NLO scale uncertainty, thus making our evaluations of the cross sections and differential distributions in $t \bar{t} Z$ production the most precise results currently  available 
in the literature.

This work completes a series of papers devoted to the study of the associated production of a top pair and a colorless heavy boson to NLO+NNLL accuracy in the partonic threshold limit.
In \cite{Broggio:2016zgg} the associated production of a top pair and a $W$ boson was studied with the methods employed here for $t \bar{t} Z$ production, while the associated production of a top pair and a Higgs boson at NLO+NNLL accuracy was considered in \cite{Broggio:2016lfj}.
In all cases the resummation was carried out in Mellin moment space. The hard and soft scales were chosen in the same way as in the traditional ``direct QCD'' approach. Codes for the numerical evaluation of the resummation are now available and tested for all of these three processes, and can be further employed in more specific phenomenological studies, according to the interests of the experimental collaborations. Within such interactions with the experimental community, a detailed study of the uncertainty associated with the choice of the PDFs and to the value of $\alpha_s(M_Z)$, in the light of a comparison with the new measurements which are expected in the forthcoming months, would be particularly illuminating. 

At this stage, it would also be interesting to combine the NLO+NNLL calculations of $t \bar{t} W$, $t \bar{t} Z$ and $t \bar{t} H$ production with the electroweak corrections for these processes \cite{Frixione:2014qaa,Frixione:2015zaa}. In addition to this, the inclusion of the decays of the heavy particles in the spirit of \cite{Broggio:2014yca} is also possible. This would allow to put kinematic cuts on the momenta of the detected particles.

\section*{Acknowledgments}
The in-house Monte Carlo code which we developed and employed to
evaluate the (differential) cross sections presented in this paper was run on the computer
cluster of the Center for Theoretical Physics at the Physics Department of New
York City College of Technology.
We thank P.~Maierhofer  and 
S.~Pozzorini for their assistance with the program {\tt Openloops}.
This research was supported by the DFG cluster of excellence ``Origin and Structure of the Universe''.
The work of A.F., G.O., and R.D.S. is supported in part by the National Science Foundation under Grant No. PHY-1417354.


\bibliography{mybib}

\bibliographystyle{JHEP}

\end{document}